\PassOptionsToPackage{sort&compress}{natbib}  
\documentclass[review]{elsarticle}

\usepackage[running]{lineno}
\modulolinenumbers[5]

\usepackage{graphicx,subfig,epsfig}
\usepackage{amssymb,amsmath,amsbsy,bm,bbm}
\usepackage{algorithm,algpseudocode,theorem}
\usepackage{natbib}
\usepackage{enumerate}   
\usepackage{multirow}
\usepackage{diagbox}
\usepackage[hyphens]{url}

\usepackage[usenames,dvipsnames]{xcolor}
\definecolor{red}{rgb}{0.7,0.0,0.0}
\definecolor{blu}{rgb}{0.0,0.0,0.7}

\def\R{\mathbb{R}}
\def\dday{\,\textrm{day}} 
\def\bc{\bar{c}}

\begin{document}

\begin{frontmatter}

\title{Modeling continuous levels of resistance to multidrug therapy in cancer}


\author[mymainaddress]{Heyrim Cho}
\ead{hcho1237@math.umd.edu}

\author[mymainaddress,mysecondaryaddress]{Doron Levy$^*$}
\cortext[mycorrespondingauthor]{Corresponding author}
\ead{dlevy@math.umd.edu}

\address[mymainaddress]{Department of Mathematics, University of Maryland, College Park, MD 20742, USA}
\address[mysecondaryaddress]{Center for Scientific Computation and Mathematical Modeling (CSCAMM), University of Maryland, College Park, MD 20742, USA}

\begin{abstract}
Multidrug resistance consists of 
a series of genetic and epigenetic alternations 
that involve multifactorial and complex processes, 
which are a challenge to successful cancer treatments. 
Accompanied by advances in biotechnology 
and high-dimensional data analysis techniques 
that are bringing in new opportunities in modeling biological systems 
with continuous phenotypic structured models, 
we study a cancer cell population model that considers a
multi-dimensional continuous resistance trait to multiple drugs 
to investigate multidrug resistance. 
We compare our continuous resistance trait model 
with classical models that assume a discrete resistance state
and classify the cases
when the continuum and discrete models yield different dynamical
patterns in the emerging heterogeneity in response to drugs.
We also compute the maximal fitness resistance trait 
for various continuum models and study the effect of 
epimutations. 
Finally, we demonstrate how our approach can be used to 
study tumor growth regarding the turnover rate and the proliferating fraction,  
and show that a continuous resistance level may result in a different
dynamics when compared with the predictions of other discrete models.
\end{abstract}


\end{frontmatter}


\section{Introduction}

The biological mechanisms responsible for the emergence of 
drug resistance and its propagation 
often involve a multifactorial and complex process of 
genetic and epigenetic alternations 
\cite{Gottesman2002,Gottesman2002a,Fodal2011}, 
that arise through a series of genetic and non-genetic changes 
\cite{Byler2014a, Byler2014b,Chaffer2011,Sarkar2013}. 
Such changes can be due to drug administration 
(drug induced resistance) {\cite{Pouchol2018,Greene2018}},  
or they can emerge independent of therapy due to intrinsic mechanisms.
Cancer cells may develop simultaneous resistance to structurally and
mechanistically unrelated drugs, leading to multidrug resistance (MDR) 
\cite{Gottesman2002,Gottesman2002a,Gottesman2010}. 
The complex dynamical nature of 
MDR is one of the most 
challenging obstacles to successful treatment.

The complexity of the mechanisms underlying drug resistance 
has encouraged its study through 
mathematical modeling.
Such models aim at providing quantitative tools for testing therapies
that circumvent or at least delay the unfortunate consequences of drug resistance.
Examples include the models of Goldie and Coldman 
\cite{Coldman1979,Coldman1983a, Coldman1983b} 
that are based on resistance due to point mutations. 
These works were proceeded by many studies considering 
stochastic models (including branching process and multiple mutations)
to study MDR and optimal control of drug scheduling 
\cite{Iwasa2006,Kimmel1998,Komarova2006,Michor2006}. 
Alternative approach includes 
continuum deterministic models using ordinary differential
equations,
for example, modeling kinetic resistance \cite{Birkhead1987} and
point mutations \cite{Tomasetti2010},
and partial differential equations, 
where spatial heterogeneity and vascularization 
can be readily incorporated \cite{Anderson1998,Tredan2007,Wu2013}. 
For additional approaches see 
\cite{LaviDoron2012,Michor2006,Foo2010,Roose2007,Coldman1998,Panagiotopoulou,Swierniak2009}.

\begin{figure}[!htb]
    \centerline{  \footnotesize (a) \hspace{5.2cm}   (b) \hspace{0.2cm}  } 
    \centerline{ 
    \includegraphics[width=5.5cm]{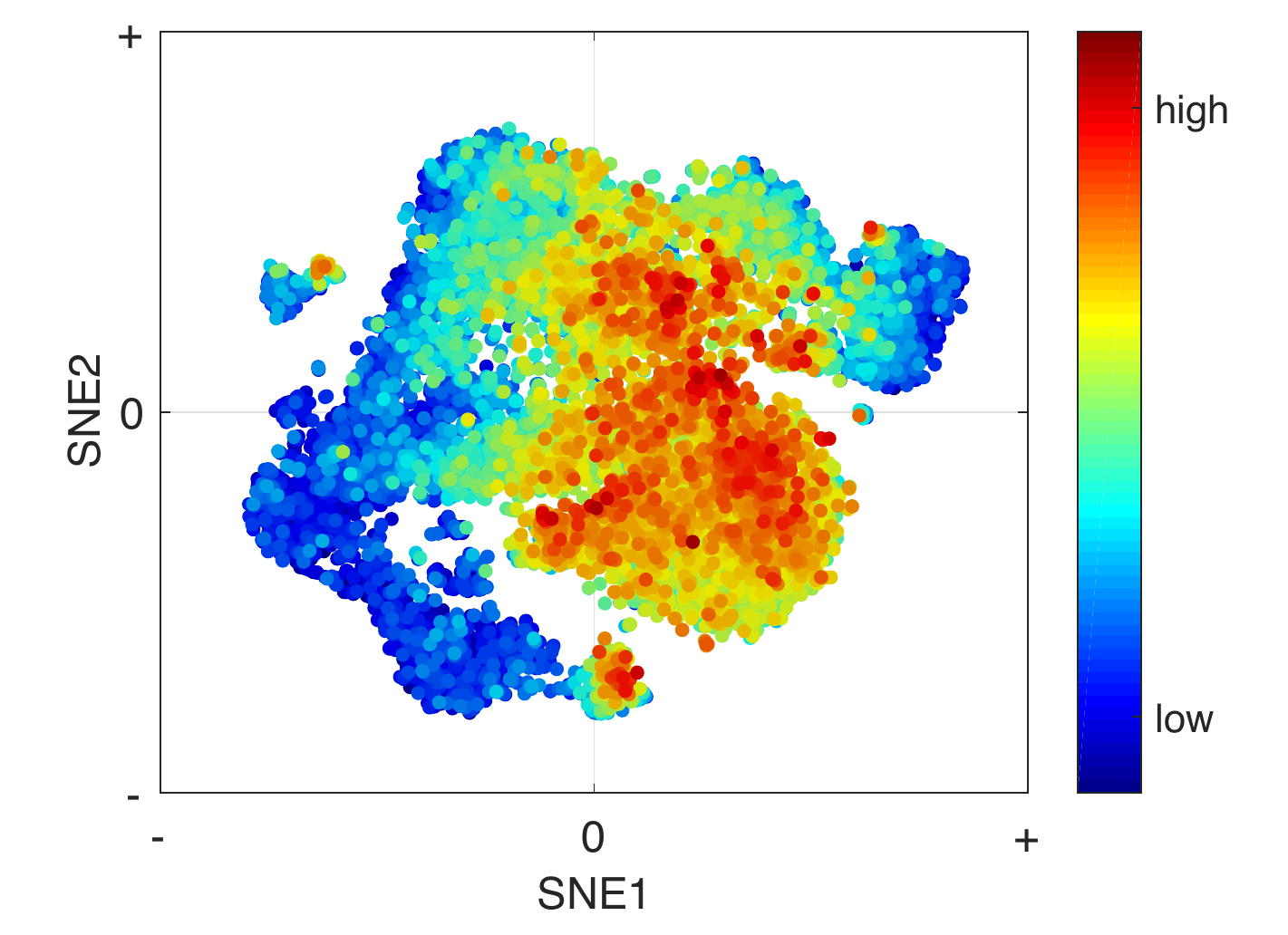} 
    \includegraphics[width=5.5cm]{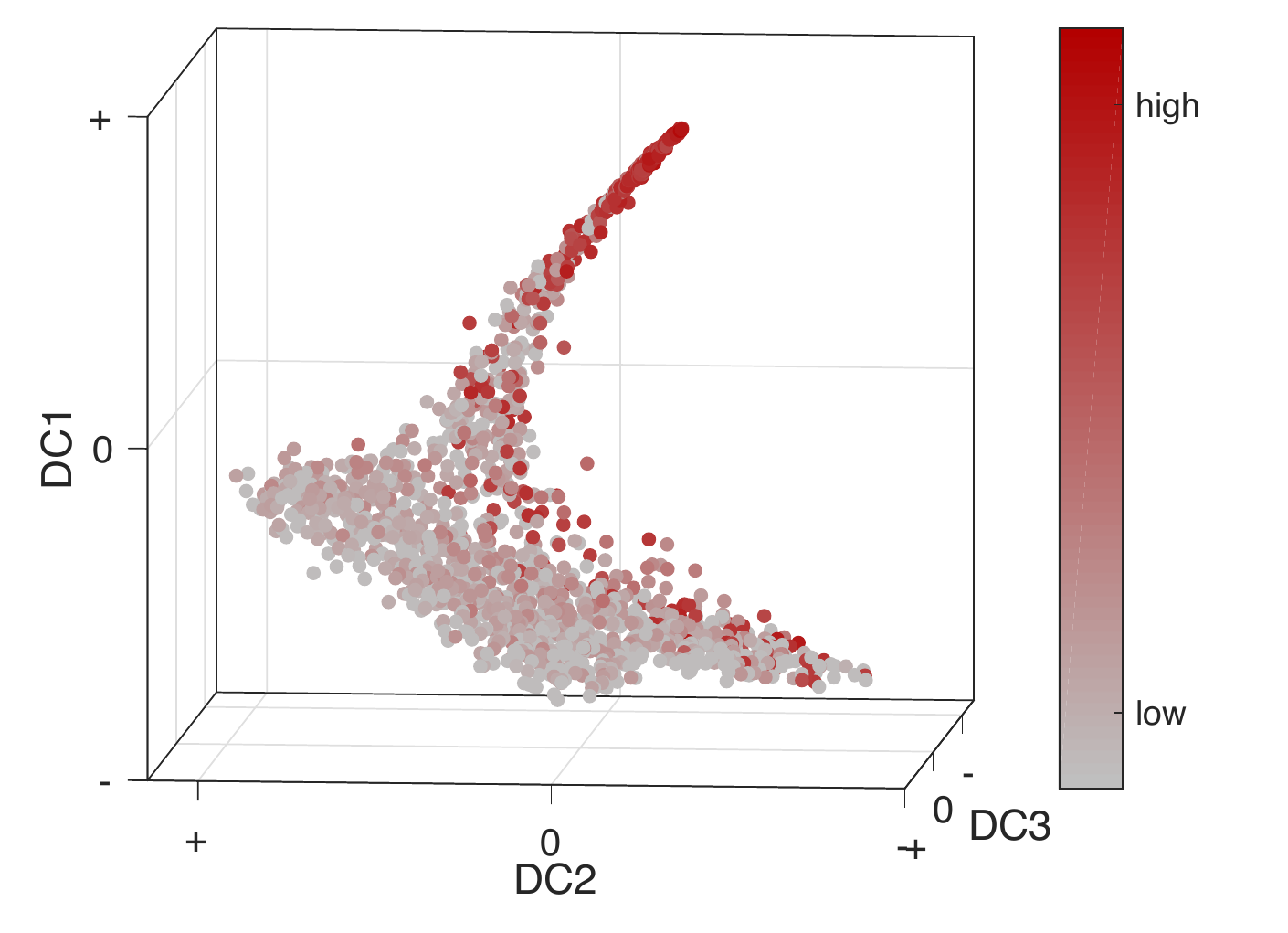} 
    }
   \caption{High-dimensional cell data 
   projected into a lower dimensional 
   continuous trait space, where the reduced dimensions are obtained by 
   dimension reduction techniques. 
   {
   Figures are reproduced from the data provided in 
   \cite{Amir2013} and \cite{Nestorowa2016}.} 
   Figure (a) shows the CD34 expression level of 
   41 dimensional data \cite{Amir2013} mapped into two dimensions by 
    stochastic neighbor embedding (viSNE), 
    and the relapsed leukemia cells are located at where CD34 is highly expressed. 
    Figure (b) 
    shows the ALDH1 expression level of 4773 dimensional data \cite{Nestorowa2016}
    mapped into three dimensions by diffusion mapping, 
    where ALDH1 is related to cancerous stem cells in mammary gland 
    and breast cancer \cite{Douville}. 
} 
\label{fig:DimReduc} 
\end{figure}

In addition to the aforementioned modeling approaches, 
the advance of biotechnology in collecting data characterizing the 
phenotype is bringing in new opportunities of mathematical modeling 
of biological systems. 
The most recent technology allows  
cytometry data to be collected up to O$(50)$ dimensions, 
Methylation profiles in the scale of O(1000), and gene-expression 
profile in the scale of O(10000) \cite{Saeys2016,Wagner2016,Macaulay2016,
Stubbington2017,Velten2017,Bendall2011}. 
In particular, 
recent advances in single cell RNA sequencing 
technologies has enabled a new high-dimensional definition of cell 
states, that is on the order of 20,000 protein encoding genes that 
compose the transcriptome \cite{Rizvi2017,Stubbington2017}. 
The high-dimensionality of the data makes it practically impossible to 
consider a meaningful model on the original space in which the data is collected. 
Thus, various dimension reduction techniques,  
such as, principal component analysis 
\cite{Mojtahedi2016,Grover2016}, 
t-distributed stochastic neighbor embedding 
\cite{Amir2013,Wagner2016,Unen2017}, 
diffusion maps \cite{Haghverdi2015,Nestorowa2016}, 
and machine learning techniques \cite{Buggenthin2017,Stumpf2017}, 
have been employed to reduce the dimensionality and to identify 
only the critical directions. 
In contrast to classical biology and modeling approaches, 
where cell types are classified into discrete states and 
differentiation is considered as a stepwise process of 
binary branching decision, 
the new technologies and data analysis 
enabled considering 
cell differentiation as a continuous process 
that can be mapped into a continuum of cellular 
and molecular phenotypes 
\cite{Macaulay2016,Haghverdi2015,Velten2017}. 
In other words, the high-dimensional configuration space 
is mapped into a {\em continuous} trait in a lower-dimensional space.
Figure \ref{fig:DimReduc} 
shows two examples of high-dimensional cell data 
mapped into a continuous trait in a lower dimensional space
using stochastic neighbor embedding (viSNE) \cite{Amir2013} 
and diffusion mapping \cite{Nestorowa2016}. 
This reveals the continuous phenotypic trait space 
where resistance can be locally characterized. 
For instance, 
the left figure shows that
relapsed leukemia cells are associated with high expression of CD34,
and the ALDH1 in the right figure is related to 
cancerous stem cells in mammary gland and breast cancer \cite{Douville}. 
This opens the door to mathematical models that assume a continuous
trait space \cite{Schiebinger,Cho2017b}.

Among continuous phenotypic structured models, 
recent studies in 
\citep{Lorz_woSp,Lorz_wSp,GreeneDoron,Cho2017,Lorenzi2016,Cho2017a} 
consider a continuous trait variable 
that represents the level of cytotoxic drug resistance. 
This framework allows to 
explicitly model the heterogeneous response to drugs
and effectively study the selection dynamics under microenvironmental constraints 
and chemotherapy. 
The asymptotic distributions on the resistance trait space 
are obtained in \cite{Lorz_woSp}, 
and the following works in \cite{GreeneDoron,Lorenzi2016} 
extend it to include mutations and epimutations. 
The distribution of resistance levels can be then translated to
therapeutic recommendation.
The effectiveness of 
a combination of cytotoxic and cytostatic drugs 
when cytotoxic resistance emerge
is studied in~\cite{Lorz_woSp}.
An optimal combination therapy 
to eliminate the most resistant clones is proposed
in~\cite{Cho2017a}.
Moreover, \cite{Cho2017a} 
extends the framework 
that was restricted to solid tumor that is 
radially symmetric with a fixed boundary 
\cite{Lorz_wSp} 
to an asymmetric tumor growth model with moving boundary. 
However, this framework is limited to 
a single trait variable to a cytotoxic drug.

In this paper we extend the framework of \cite{Cho2017a} to  
multi-dimensional resistance trait.  We compare our approach that
allows for a continuous drug response to more traditional
approaches that assume a discrete response to drugs.
The paper is organized as follows. 
In section \ref{sec:meth}, we introduce a mathematical model 
for MDR assuming continuous trait variables.
We parameterize our model as an extension of 
a discrete resistance state model in section \ref{sec:2-1} 
and compute the maximal fitness trait of resistance in 
section \ref{sec:2-2} for different types of 
continuum models. This allows us 
to characterize the cases when  
the solutions of the continuous models are qualitatively different
than the corresponding discrete models.
Section \ref{sec:2-3} presents simulation results 
for the different cases of cytotoxic and cytostatic drugs studied in~\ref{sec:2-2}.
The impact of mutations and epimutations 
is studied in section~\ref{sec:2-4}.  
In section~\ref{sec:Num} we simulate tumor growth and resistance dynamics 
subject to MDR on different types of tumors 
characterized by turnover rates and the proliferating ratios.
Our simulations correspond to the discrete MDR models studied by
Komarova and Wodarz (2005)  \cite{komarova2005} 
and Gardner (2002)  \cite{gardner2002}. 
We observe that a combination therapy 
with multiple cytotoxic drugs is also effective in 
high turnover tumors using relatively high dosages. 
Increasing the dosage in 
low turnover tumor is effective only for 
certain drug uptake functions. 
In addition, 
the drug response function plays a key role in
determining the tumor growth dynamics when
combination therapy is administered using  
cell-cycle nonspecific cytotoxic drugs, 
such as Cyclophosphamide and Doxorubicin. 
Conclusions and future directions are discussed in section~\ref{sec:Summary}.

\section{Models of multidrug resistance}\label{sec:meth}

Let us consider a cancer growth model under multidrug therapy that 
depends on an $M$-dimensional phenotype variable 
$\theta = ( \theta_1,\, ...,\, \theta_M ) \in \Gamma \doteq \Pi_{i=1}^M \Gamma_i$.
The phenotype variable in the $i$-th direction 
$\theta_i \in \Gamma_i = [0,\,1]$ characterizes  
the resistance level to the $i$-th drug or the $i$-th 
drug mechanism, 
where $\theta_i = 0$ and $\theta_i = 1$ 
represents the fully-sensitive cells and fully-resistant cells 
to drug $i$, respectively. 
The value of $\theta_i$ can be obtained by 
normalizing 
the expression level of a gene or a gene cluster 
that is linked to the cellular 
levels of drug resistance and proliferative
potential, such as ALDH1, CD44, CD117, or MDR1 
\cite{Amir2013,Hanahan2011,Medema2013,Pisco2015}.  
The governing equations follows the dynamics of 
the density of proliferating cells, $n_P = n_P(t,\theta)$,
and quiescent cells, $n_Q = n_Q(t,\theta)$, as 
\begin{eqnarray}
	\partial_t n_P(t,\theta) &=&  \left( (1-w)R(t,\theta) - D - C_P(t,\theta) - q
	 \right) n_P(t,\theta) \label{eq:Gvrn0} \\ \nonumber 
	 & & + p n_Q(t,\theta) + w \int_\Gamma M(\theta, \vartheta) R(t,\vartheta) n_P(t,\vartheta) d \vartheta, \\ 
	\partial_t n_Q(t,\theta) &=&  q n_P(t,\theta) + \left( -p - D_Q - C_Q(t,\theta) 
	 \right) n_Q(t,\theta).  \label{eq:Gvrn1} 	 
\end{eqnarray} 
The first term on the RHS of Eq.~(\ref{eq:Gvrn0}) is a growth term,
$R(t,\theta)$, which we assume
depends on 
the resource level $s_0(t)$ with the proliferation rate function 
$\varphi(\theta)$ as 
$R(t,\theta) = \varphi(\theta) s_0(t)$. 
Also, we assume an exponential growth 
by considering a constant apoptosis rate $D$ for 
the proliferating cells and $D_Q$ for the quiescent cells. 
To consider a logistic growth, 
we  substitute both terms with  
a density-dependent apoptosis term $d \rho(t)$,
where $\rho(t)$ is the total number of cells 
\begin{equation*}
\rho(t) = \int_\Gamma n_P(t,\theta) + n_Q(t,\theta) d\theta,
\end{equation*}
and $d$ is a constant that determines the cell capacity.  

The net effects of the cytotoxic drugs on the proliferating and
quiescent cells are denoted as 
$C_P(t,\theta)$ and $C_Q(t,\theta)$, respectively.
These terms depend 
on the marginal drug effects, $C_i = C_i(t,\theta; c_i(t))$, the cell death rate 
due to the $i$-th drug, which is assumed to be 
a function of the drug concentration $c_i(t)$. 
We either consider 
$C_i(t,\theta) = \mu_i(\theta) c_i(t)$,  
where $\mu_i(\theta)$ is the drug uptake function of the $i$-th drug,  
or the exponential kill model \cite{gardner2002}, $C_{i}(\theta_i) = e^{-a_i (\theta_{max}-\theta_i) c_i(t) }$, where 
$C_i$ represents the probability of the cell death 
due to the $i$-th drug.  
The net drug effect is modeled as 
$C_P(t,\theta) = \Phi( C_1, ..., C_M)$, where 
$\Phi$ is the overall drug effect function 
that can be taken 
for the cytotoxic drugs as 
\begin{equation}
C_P(t,\theta) = \Phi(C_1, ..., C_M) = 1 - \prod_i (1-C_i),
\label{eq:c-star}
\end{equation}
and similarly for $C_Q$. 
The form (\ref{eq:c-star}) is valid  
when $C_i$ is the probability of death due to the $i$-th drug ($C_i
\leq 1$), and
assuming that the drug effects are independent. 
Dependency between the drugs can be imposed through 
different choices of $\Phi$, e.g., Copula functions \cite{Copula}
that are used to describe the dependence between random variables 
using multivariate probability distributions 
with prescribed marginal distribution functions. 
In addition to the cytotoxic drugs, we consider cytostatic drugs,
which we assume delay the proliferation according to
\begin{equation*}
R(t,\theta) = \dfrac{\varphi(\theta) s_0(t)}{1 + \Phi(C_1, ..., C_M) }.
\end{equation*}
The net cytostatic drug effect delays the progression 
of the proliferating cells 
through the cell cycle.
We assume an additive $\Phi$:
\begin{equation*}
\Phi(C_1, ..., C_M) = \sum_i C_i.
\end{equation*}

Proliferating cells enter the quiescent state at a rate $q$ and 
quiescent cells return to the cycling compartment at a rate $p$. 
These rates regulate 
the proliferating portion $\delta(t) \doteq \int n_P d\theta / \rho(t)$. 
To balance a fixed ratio of proliferating cells, namely 
the proliferating index $\delta^*$, 
the transfer rate $q$ can be computed as 
$q = (\max_{\theta}R(\theta)-D+D_Q) (1-\delta^*) 
+ p{(1-\delta^*)}/{\delta^*}.$

The last term in Eq.~(\ref{eq:Gvrn0}) is a mutation term.  
We assume that mutations  occur at a rate 
$w$ during the proliferation cycle.
The mutation is modeled as a integral term with a 
kernel function $M(\theta, \vartheta)$. 
$M(\theta, \vartheta)$ represents the probability 
of a mother trait $\vartheta$ mutating to a daughter trait 
$\theta$ that is taken as an 
asymmetric exponential function with mutation range $\ell$, i.e., 
$M(\theta, \vartheta) = M_0 \exp\left[ (\theta - \vartheta )^2 
/ \ell^2 \right]$ 
for $\theta  \geq \vartheta$, and zero otherwise. 
Here, $M_0$ is a normalizing constant. 
This model represents a mutation 
that gradually increases the resistance level through multiple mutations. 
A rare mutation that 
confers a complete drug resistance in a single step can be 
imposed with a discrete kernel function \cite{Cho2017} and 
a smaller value of $w$.

\subsection{Multidrug resistance models parameterized with a binary level of resistance}\label{sec:2-1}

In this section, we simplify the model given by Eq.~\eqref{eq:Gvrn0} 
to a model that assumes a binary trait space. In this case, cells
are either fully-sensitive or fully resistant with respect to each
drug, i.e., $\theta_{i} \in \{0,1\}, \forall i$.
To compare the discrete- and continuous-trait models, 
we parameterize the proliferation and drug function with 
the parameters related to the microenvironment selection 
as follows. 
We denote the proliferation rate 
of the fully-sensitive cells ($\theta = 0$) as $\gamma$, 
and assume that 
the proliferation rate of the fully-resistant cells ($\theta = 1$) 
is reduced by $\eta$. 
With a normalized constant resource level ($s_0=1$), 
\begin{equation*}
R(0) = \varphi(0) = \gamma, \qquad R(1) = \varphi(1) = \gamma-\eta.
\end{equation*}
We scale the drug dosage $c(t)$ 
to represent the drug effect on the fully-sensitive cells 
and assume that the fully-resistant cells do not respond to 
the drug. This yields a drug uptake function for which  
$\mu(0) = 1$ and $\mu(1) = 0$.
Hence, the effect of the cytotoxic drug 
$C(t, \theta) = c(t) \mu(\theta)$  boils down to
\begin{equation*}
C(t, 0) =  c(t), \qquad C(t, 1) = 0.
\end{equation*}
See Table \ref{Tbl:param} for a summary of the fitness parameters. 
\begin{table}[htb!] 
\center
\begin{tabular}{c|lll} \hline 
parameters & biological meaning \\ \hline 
$\gamma$ & maximum proliferation rate  \\ 
$\eta$ & reduced proliferation due to resistance (selection gradient) \\ 
$c(t)$ & maximum apoptosis rate of sensitive cells due to drug \\  \hline 
\end{tabular}
\caption{Parameters of the proliferation and drug effect 
that yield the microenvironmental selection process \cite{Lorenzi2016}.  }
\label{Tbl:param}
\end{table}

The resulting model can be written as a dynamical system.
For instance, we consider a single ($M=1$) cytotoxic 
drug affecting the proliferating cells.
There exists two cell states:
sensitive cells, $n_S(t) \doteq n_P(t,\theta=0)$, and resistant cells,
$n_R(t) \doteq n_P(t,\theta=1)$.
In this case, the resulting system is
\begin{equation}
\begin{aligned}
	\dot{n_S} &=& \left( (1-w)\gamma - D - c(t) \right)  n_S, \\	
	\dot{n_R} &=& w \gamma n_S + \left( (\gamma-\eta) - D \right)  n_R,  
\end{aligned}  
\label{eq:GvrnDiscrete}
\end{equation}
where, $D=d\rho(t)$, and $\rho(t) = n_S(t) + n_R(t)$. 
In the case of a single cytostatic drug affecting the proliferating
cells, the dynamics follows
\begin{equation}
\begin{aligned}
	\dot{n_S}  &= \left( (1-w)\frac{\gamma}{1+c(t)} - D \right)  n_S, \\	
	\dot{n_R}  &= w \frac{\gamma}{1+c(t)}n_S + \left( (\gamma-\eta) - D \right)  n_R.  
\end{aligned}  
\label{eq:GvrnDiscrete_}
\end{equation}
In case of $M$ drugs, the resulting model will involve
$2^M$ discrete cell state variables. 

The binary models \eqref{eq:GvrnDiscrete} 
and \eqref{eq:GvrnDiscrete_} 
yield an outcome where either the sensitive cells $n_S$ or the resistant 
$n_R$ cells dominate the population asymptotically 
depending on the fitness parameters. 
In particular, for Eq.~\eqref{eq:GvrnDiscrete}, 
with fixed values of $\gamma$ and $\eta$, 
if the drug dosage is low, $c(t) < \eta-w\gamma$,
the sensitive cells dominate, 
but if the drug dosage increases as $c(t) \geq \eta-w\gamma$, 
the resistant cells dominate the population. 
The same holds for Eq.~\eqref{eq:GvrnDiscrete_} 
with a threshold 
$(\eta - w \gamma) / (\gamma-\eta)$.
If the mutation during treatment is negligible ($w = 0$) \cite{komarova2005}, 
the thresholds become $\eta$ and $\eta / (\gamma-\eta)$ for models
(\ref{eq:GvrnDiscrete}) and (\ref{eq:GvrnDiscrete_}), respectively.

To connect between models with binary traits and models with
continuous traits, we extend the binary models assuming
that the proliferation and drug effects are smooth and monotone
with respect to $\theta$. 
This assumption (although may not always hold) 
makes it possible to 
classify continuum scenarios 
and helps in identifying cases in which the continuous
traits dynamics is qualitatively different than the
corresponding binary models.
Since we only consider proliferating cells, the transfer terms 
to the quiescent cells are removed from Eq.~\eqref{eq:Gvrn0}, 
and we simulate 
\begin{eqnarray}
	\partial_t n(t,\theta) &=&  \left( R(\theta) - D - C(t,\theta) 
	 \right) n(t,\theta) \label{eq:Gvrn2}. 
\end{eqnarray} 

\begin{figure}[!htb]
    \centerline{ 
    \includegraphics[width=10cm]{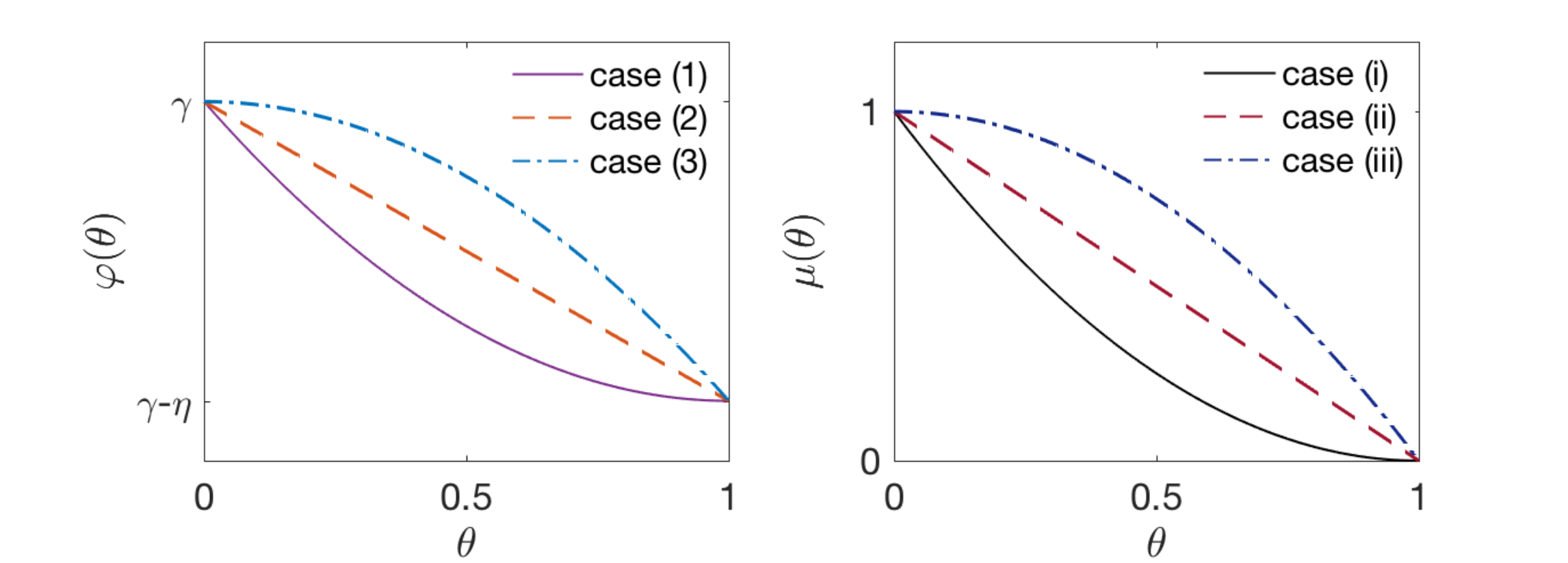} 
    }
   \caption{ Models of proliferation rate $\varphi(\theta)$ and 
   drug uptake $\mu(\theta)$ 
   considering a continuous resistance trait space on $\theta \in [0,\,1]$. We assume that the proliferation rate reduces from $\gamma$ to $\gamma-\eta$ as the resistance level increases, and the drug effect reduces from1 to 0. } 
\label{fig:Conti_RnC}	
\end{figure}

Starting from the proliferation, we assume that 
cells that are resistant to cytotoxic drugs use their 
resources to develop and maintain the drug resistance mechanism 
\citep{Mumenthaler2015,Wosikowski}, 
that is, $\varphi'(\theta) < 0$. 
On the domain of $\theta \in [0,\,1]$, 
the proliferation function $R(\theta) = \varphi(\theta)$ 
can be characterized according to its concavity.
We consider three sample cases:
$\varphi(\theta) = \gamma - \eta + \eta (\theta-1)^2$, 
$\varphi(\theta) = \gamma - \eta \theta$, 
and $\varphi(\theta) = \gamma - \eta \theta^2$. 
The cytotoxic drug effect $C(\theta) =  c(t) \mu(\theta)$ 
can be modeled similarly. 
Assuming that apoptosis decreases with an increased level of resistance,
we have 
$\mu'(\theta) < 0$. 
Accordingly, we consider three characteristic cases:
$\mu(\theta) = (\theta-1)^2$, 
$\mu(\theta) = ( 1  -  \theta)$, and 
$\mu(\theta) = ( 1  -  \theta^2)$. 
The models we consider are summarized in Table \ref{Tbl:caseRnC} 
and Figure \ref{fig:Conti_RnC}. 
\begin{table}[htb!] 
\center
\begin{tabular}{c|ccc} \hline\hline  
 &   concave up  & 
  $\hspace{.7cm}$ linear  $\hspace{.9cm}$  & $\hspace{.3cm}$ concave down  $\hspace{.3cm}$ \\ \hline 
$\varphi(\theta)$ & 
(1) $\gamma - \eta + \eta (\theta-1)^2$ & 
(2) $\gamma - \eta \theta$ &  
(3) $\gamma - \eta \theta^2$ \\ 
$\mu(\theta)$ & 
(i) $(\theta-1)^2$ &  
(ii)  $( 1  -  \theta)$ & 
(iii)  $ 1  -  \theta^2$ \\ \hline \hline  
\end{tabular}
\caption{Classification of the continuous proliferation and drug effect 
functions depending on the concavity. We consider three cases for both $R(\theta) = \varphi(\theta)$ and $C(\theta) = \mu(\theta) c(t)$ 
denoted as case $\{ 1,\, 2,\, 3 \}$ and 
$\{$i, ii, iii$\}$, respectively.}
\label{Tbl:caseRnC}
\end{table}

\subsection{Differentiating models with binary traits from models
  with continuous traits}  \label{sec:2-2}

To demonstrate the difference between models that are based on
binary traits and continuous-traits models, 
we compute the trait that achieves 
the maximal fitness of Eq.~\eqref{eq:Gvrn2} 
under different microenvironment conditions. 
We denote such trait with the maximal 
growth rate as 
$\theta_M( c(t), \eta, \gamma ) \doteq \arg\max_{\theta} \left( R(\theta)-C(\theta) \right)$.  
Our choices of $R(\theta)$ and $C(\theta)$ in Section~\ref{sec:2-1} yield 
nine cases that are presented in the following 
list\footnote{For simplicity, we compute the maximal fitness trait 
following the assumption that 
mutations during treatment are negligible ($w = 0$) \cite{komarova2005}.}. 
We comment that among the nine cases, 
six cases resemble the discrete model in a sense that 
the maximal fitness trait is binary, either fully-sensitive 
or fully-resistant, while three cases allow 
intermediate trait levels. 
This demonstrates that in certain circumstances, 
continuum models are necessary. 
We first consider the single cytotoxic drug setup that is 
comparable to the binary model~\eqref{eq:GvrnDiscrete}.  
The results are summarized in Table~\ref{Tbl:maxfit}.
\begin{itemize}
\item 
Case (3,i). The maximal growth rate is achieved at $\theta_M = {c(t)}/{( \eta+c(t) )}$ that changes its value from 
$\theta_M(c=0,\cdot,\cdot) = 0$ to $\lim_{c \rightarrow \infty} \theta_M (c,\cdot,\cdot) = 1$. 
This case 
allows an intermediate maximal fitness trait for any drug dosage 
$c(t) \in \R_+$. 
\item 
Case (3,ii). The maximal growth rate is achieved at $\theta_M =
{c(t)}/{( 2 \eta )}$.
This model increases the maximal trait linearly in terms of the drug dosage when $c(t) \leq 2\eta$. For $ c(t) > 2\eta $, the maximal fitness occurs at $\theta_M=1$. 

\item 
Case (3,iii). The maximal growth rate is either achieved at 
$\theta_M=0$ when $ c(t) < \eta $, or at 
$\theta_M=1$ when $ c(t) > \eta $. 
Since the phenotype distribution asymptotically converges to 
a delta function centered at 
$\theta=0$ or 
$\theta=1$, the overall quality of the solution is 
similar to the binary-trait model. 
We also remark that there exists a critical drug dosage 
at $ c(t) = \eta $ that yields multiple fitness traits. 

\item 
Case (2,i). This model is similar to the case (3,ii), but opposite 
in the sense that the maximal growth rate is achieved at 
$\theta_M=0$ for $c(t) <\eta/2$, and increases as 
$\theta_M = {(2c(t)-\eta)}/{ 2 c(t) }$ for $c(t) \geq \eta/2$. 

\item 
Cases (2,ii), (2,iii), (1,i), (1,ii), and (1,iii). 
These models also yield a solution that is either 
concentrated at $\theta_M=0$ or $\theta_M=1$, 
similar to case (3,iii), that is, 
$\theta_M = \mathbf{1}_{c > \eta}$, 
where $\mathbf{1}_A$ is an indicator function on $A$.  
\end{itemize} 

\begin{table}[htb!] 
\center 
\begin{tabular}{c|c|c|c} \hline \hline 
   \diagbox{$\mu$}{$\varphi$} &  $\quad$ Case (1) $\quad$  & $\qquad$ Case (2)  $\qquad$  & $\quad$  Case (3) $\quad$   \\ \hline 
(i) & 
 $\theta_M = \mathbf{1}_{c > \eta}$   & 
  $\theta_M = \max\left( 0, 
   \dfrac{2c-\eta}{2 c} \right) $ &  
  $\theta_M = \dfrac{c}{ \eta+c }$ \\ [9pt] \hline 
(ii) & 
 $\theta_M = \mathbf{1}_{c > \eta}$  &  
 $\theta_M = \mathbf{1}_{c > \eta}$  & 
  $\theta_M = \min \left(  
  \dfrac{c}{ 2 \eta }, 1  \right) $ 
 \\ [9pt]  \hline  
(iii) & 
 $\theta_M = \mathbf{1}_{c > \eta}$   & 
  $\theta_M = \mathbf{1}_{c > \eta}$  & 
 $\theta_M = \mathbf{1}_{c > \eta}$  
 \\ [6pt] \hline \hline 
\end{tabular}
\caption{The selected trait with maximal growth rate 
$\theta_M = \theta_M( c, \eta, \gamma)$ depending on 
the {\em cytotoxic} drug concentration $c$ and 
the resource parameters $\gamma$ and $\eta$.
}
\label{Tbl:maxfit}
\end{table}

In addition to cytotoxic drugs, we also consider 
the drug uptake models in Table~\ref{Tbl:caseRnC} 
for a single cytostatic drug that is comparable to 
the binary model~\eqref{eq:GvrnDiscrete_}. 
The maximal fitness traits for the different choices of proliferation
rate functions and drug uptake functions are summarized in
Table~\ref{Tbl:maxfit2}.

\begin{table}[htb!] 
\center 
\begin{tabular}{c|c|c|c} \hline \hline 
 \diagbox{$\mu$}{$\varphi$} &  $\quad$ Case (1) $\quad$  & $\qquad$ Case (2)  $\qquad$  & $\quad$  Case (3) $\quad$   \\ \hline 
\multirow{2}{*}{(i)} & 
 \multirow{2}{*}{
$\theta_M = \mathbf{1}_{c > \frac{\eta}{\gamma-\eta}}$ 
 }
 & 
$\theta_M  = \frac{\gamma}{\eta} - \sqrt{\frac{\gamma^2}{\eta^2} - C_{2i}} $,  
 &  
$\theta_M  =  \frac{C_{1i}}{2} - \sqrt{\frac{C_{1i}^2}{4} - \frac{\gamma}{\eta} }$, 
 \\ 
 & 
 & 
where $C_{2i} = \frac{2\gamma}{\eta}-\frac{1}{c}-1 $ 
 &
where $C_{1i} = 1+\frac{1}{c} + \frac{\gamma}{\eta}$ 
 \\ [0pt] \hline  
\multirow{2}{*}{(ii)}  & 
 \multirow{2}{*}{  
$\theta_M = \mathbf{1}_{c > \frac{\eta}{\gamma-\eta}}$ 
 }  
&
\multirow{2}{*}{  
$\theta_M = \mathbf{1}_{c > \frac{\eta}{\gamma-\eta}}$ 
 }
 & 
 $\theta_M = C_{1ii} - \sqrt{C_{1ii}^2 - \frac{\gamma}{\eta} }$, 
 \\ 
 & 
 &  
 & 
 where $C_{1ii} = 1+\frac{1}{c}$ 
  \\ [0pt]  \hline  
(iii) & 
$\theta_M = \mathbf{1}_{c > \frac{\eta}{\gamma-\eta}}$ 
  & 
$\theta_M = \mathbf{1}_{c > \frac{\eta}{\gamma-\eta}}$  & 
$\theta_M = \mathbf{1}_{c > \frac{\eta}{\gamma-\eta}}$   \\ [5pt] \hline \hline 
\end{tabular}
\caption{The selected trait with maximal growth rate 
$\theta_M = \theta_M( c, \eta, \gamma)$ depending on 
the {\em cytostatic} drug concentration $c$ and 
the resource parameters $\gamma$ and $\eta$. We remark that 
$\theta_M$ are taken as 0 or 1 in 
cases (2,i) and (3,ii) 
similar to Table \ref{Tbl:maxfit}. 
}
\label{Tbl:maxfit2}
\end{table}

\subsection{Simulation of continuum model in cytotoxic and cytostatic resistance} \label{sec:2-3}

In this section, we simulate the model \eqref{eq:Gvrn2} 
for the cases shown
in Table \ref{Tbl:caseRnC} 
and compare the results with 
the binary models 
\eqref{eq:GvrnDiscrete}--\eqref{eq:GvrnDiscrete_}. 
For the numerical simulations, we consider the 
maximal proliferation rate as $\gamma = 0.66$ per day, 
corresponding to a cell cycle of approximately 25 hours \cite{Steel1966,Calabresi}.
We also assume that the reduction in proliferation of
the resistant cells is 
$\eta = 0.132$ per day 
based on 
the experiments of non-small lung cancer cells exposed to Erlotinib \cite{Mumenthaler2015}, 
where the 
growth rate of resistant cell is reduced by approximately 70\%. 
Experiments with HL60 leukemic cells exposed to vincristine \cite{Pisco2013} and calculation in \cite{Lorenzi2016} further support this assumption. 
We assume a logistic growth by 
$D = d\rho(t)$, where 
the apoptosis constant that represents 
the average death rate is taken as $d = 0.66 \cdot 10^{-8}$. 
This corresponds to a cell capacity of $10^8$ \cite{Pisco2013} 
assuming a solid tumor of size 1cm$^3$ 
prior to angiogenesis \cite{Anderson2005} and 
a tumor cell volume $10^{-9}\sim 3\cdot 10^{-8}$cm$^3$ 
\cite{Casciari,Folkman}.

\begin{figure}[!htb]
    \centerline{ 
    \includegraphics[width=11cm]{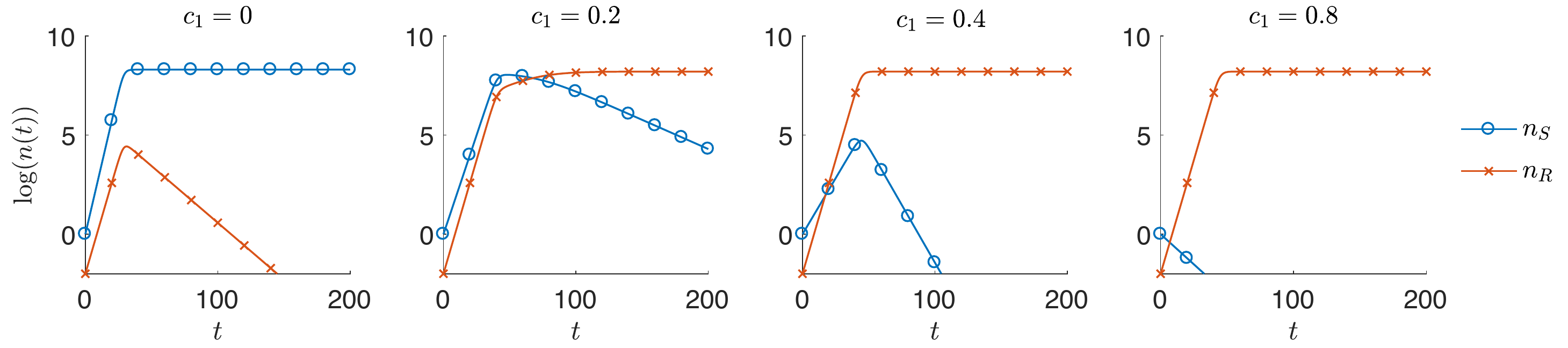} 
    }
   \caption{Total number of sensitive and resistant cancer cells in
     log scale using the binary-trait model \eqref{eq:GvrnDiscrete} for
     different dosages of cytotoxic drug. The outcome is
     asymptotically binary, where either the sensitive or resistant
     cells dominate depending on the drug dosage with a threshold $c_1
     = \eta = 0.132$. 
     }
\label{fig:nPntp_bar_c1} 
\end{figure} 

\begin{figure}[!htb] 
    \centerline{ \footnotesize\rotatebox{90}{\hspace{0.7cm} case (3,i) } 
    \includegraphics[width=11cm]{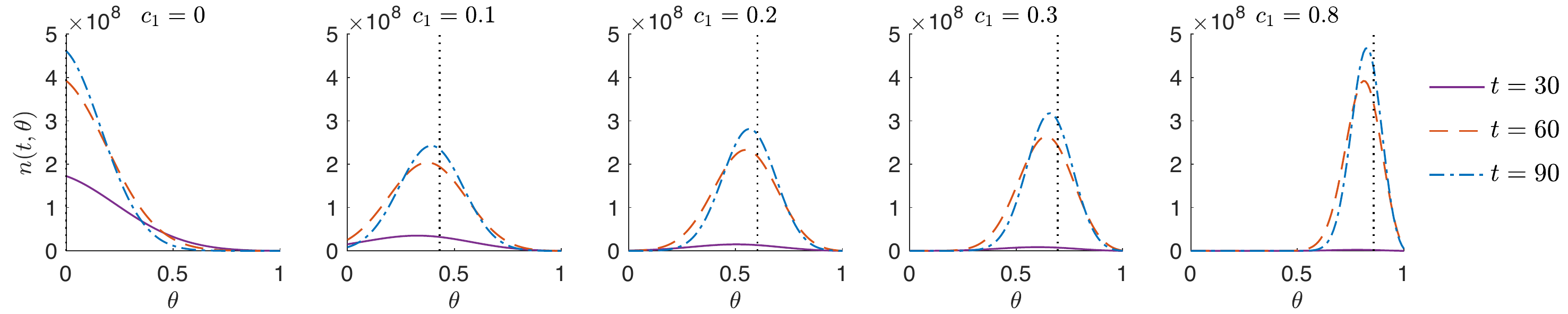} 
        }
    \centerline{ \footnotesize\rotatebox{90}{\hspace{0.6cm} case (3,ii)} 
    \includegraphics[width=11cm]{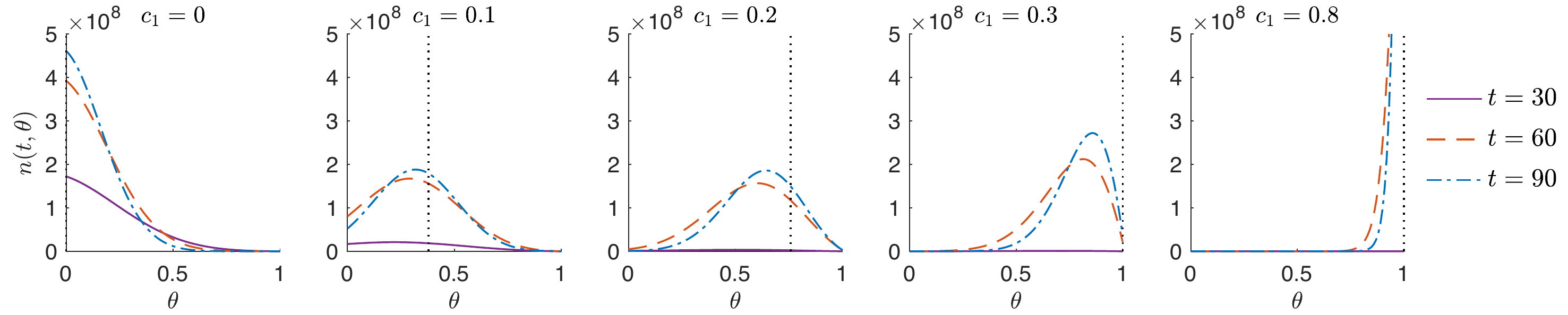} 
         }
    \centerline{ \footnotesize\rotatebox{90}{\hspace{0.7cm} case (2,i) } 
    \includegraphics[width=11cm]{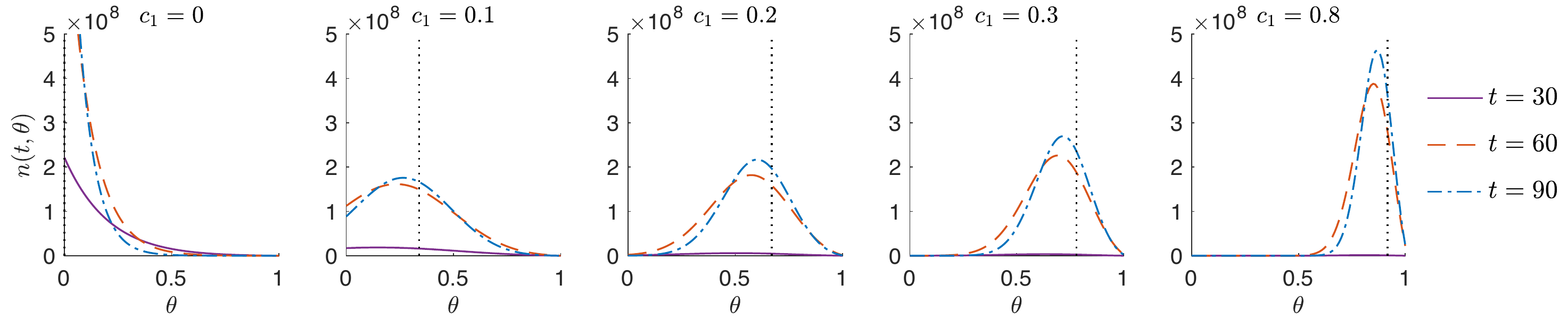} 
        }
    \centerline{ \footnotesize\rotatebox{90}{\hspace{0.6cm} case (2,ii) } 
    \includegraphics[width=11cm]{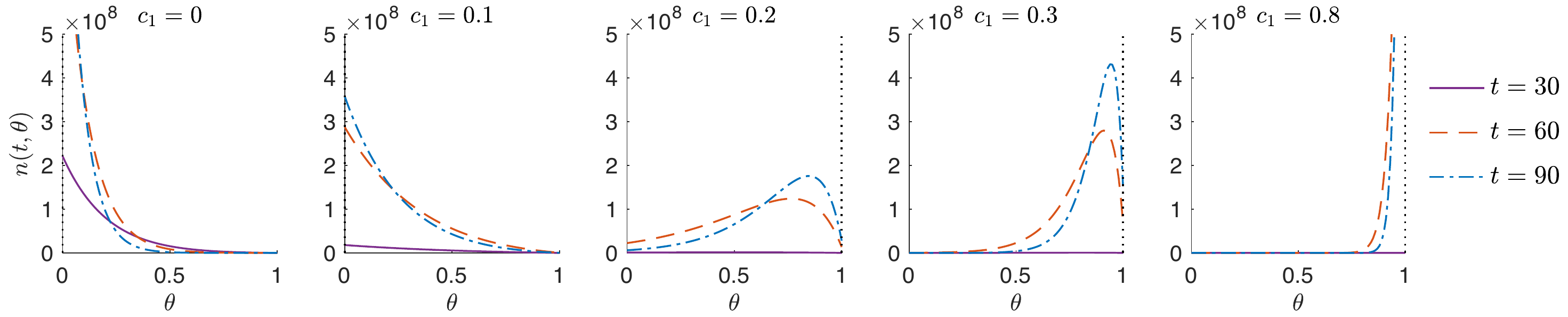} 
        }
    \centerline{ \footnotesize\rotatebox{90}{\hspace{0.5cm} case (1,iii) } 
    \includegraphics[width=11cm]{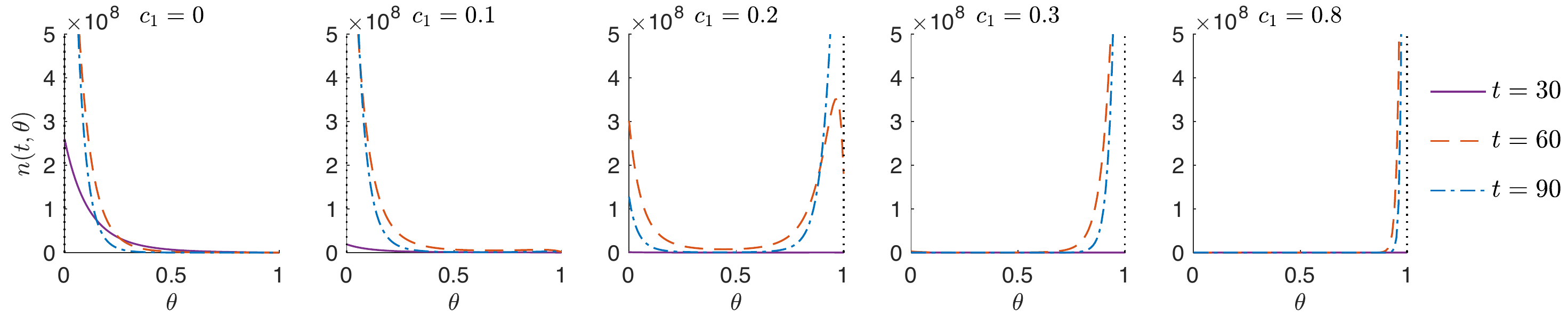} 
        }  
   \caption{The dynamics of the resistance profile of the cancer cells
     in the continuous-trait model \eqref{eq:Gvrn2}. 
   The drug dosages are considered from $c_1 = 0$ to $0.8$ 
   and the shown results are at time $t = 30$, 60, and 90. 
   Cases (3,i), (3,ii), and (2,i) yield a distribution 
   with an intermediate resistance level of maximal fitness, 
   where the maximum trait occurs at 
   $\theta_M(c_1) = \frac{c_1}{ 0.132+c_1 }$,  
   $\theta_M(c_1) = \frac{c_1}{ 0.264 }$, and 
   $\theta_M(c_1) = \frac{2 c_1- 0.132}{2 c_1 }$, respectively. 
   Cases (2,ii) and (1,iii) result in a 
   distribution that is similar to the binary-trait model, 
   either concentrated at the fully sensitive 
   or fully resistant trait (see Table \ref{Tbl:maxfit}). 
   } 
\label{fig:Pntp_allc1}	
\end{figure}

In Figure~\ref{fig:nPntp_bar_c1}, 
we first present the result of the binary-trait model 
\eqref{eq:GvrnDiscrete} 
showing that either the fully-resistant or the fully-sensitive cells 
survive depending on the drug dosage 
$c_1(t)$ compared to $\eta = 0.132$. 
The total number of sensitive and 
resistant cells, $n_S(t)$ and $n_R(t)$, are plotted in log scale with 
a constant drug dosage up to time $t=200$. 
We observe that when $c_1 = 0 < \eta$, the sensitive cells dominate at 
$t = 200$, however, when the drug dosage increases to $c_1 \geq 0.4 > \eta$, 
the resistant cells dominate. 
When $c_1 = 0.2 > \eta$, but close to $\eta$,  
the resistant cells will eventually dominate.  

In contrast, Figure \ref{fig:Pntp_allc1} shows 
the cancer cell density $n(t,\theta)$ of 
the continuous-trait model  \eqref{eq:Gvrn2} 
subject to cytotoxic drug 
for cases (3,i), (3,ii), (2,i), (2,ii), and (1,iii). 
We vary the constant cytotoxic drug dosage 
from $c_1 = 0$ to $0.8$ and compute the solution up to time $t=90$. 
Case (3,i) always yields an intermediate level 
of maximal fitness trait 
of resistance level  
$\theta_M(c_1) = \frac{c_1}{\eta+c_1 }$. 
Case (3,ii) also yields intermediate levels 
of $\theta_M(c_1) = \frac{c_1}{ 2 \eta}$ when 
$c_1 \leq 2\eta =  0.264$, and $\theta_M(c_1)=1$ otherwise. 
Alternatively in case (2,i), 
$\theta_M(c_1)=0$ when  $c_1 < \eta/2 = 0.066$, 
and 
$\theta_M(c_1) = \frac{2 c_1- \eta}{ 2 c_1 }$ otherwise.  
These simulations are consistent with
Table~\ref{Tbl:maxfit}.  
Moreover, 
we observe that the transition from the sensitive to 
the resistant trait is faster 
in cases (3,ii) and (2,i) compared with case (3,i), 
and even more rapid in cases (2,ii) and (1,iii). 
In particular, cases (2,ii) and (1,iii) result in 
a distribution that is either concentrated 
at the fully sensitive or fully resistant trait with a 
threshold $c_1 = \eta = 0.132$.  

\begin{figure}[!htb] 
    \centerline{  \footnotesize  (a) binary \hspace{2.1cm}   (b) case (3,i)  \hspace{1.9cm}  (c) case (1,iii) \hspace{0.3cm} } 
    \centerline{ 
    \includegraphics[width=3.5cm]{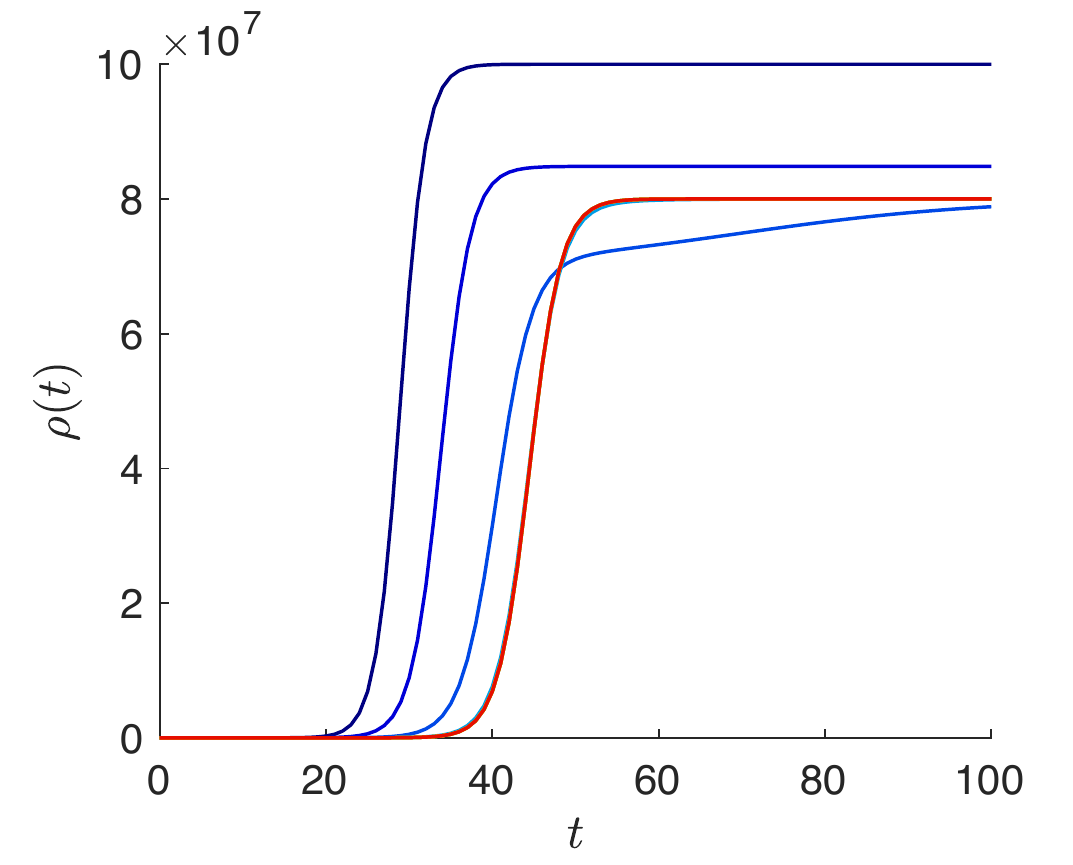} 
    \includegraphics[width=3.5cm]{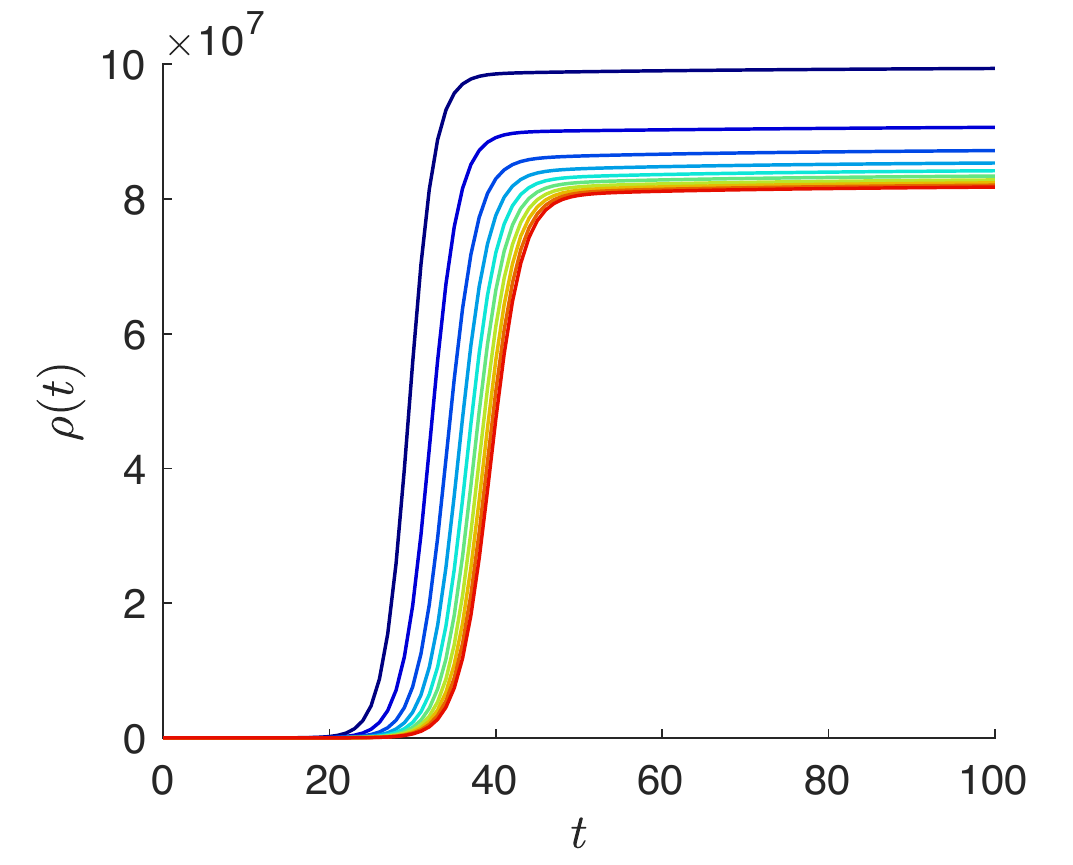} 
    \includegraphics[width=3.5cm]{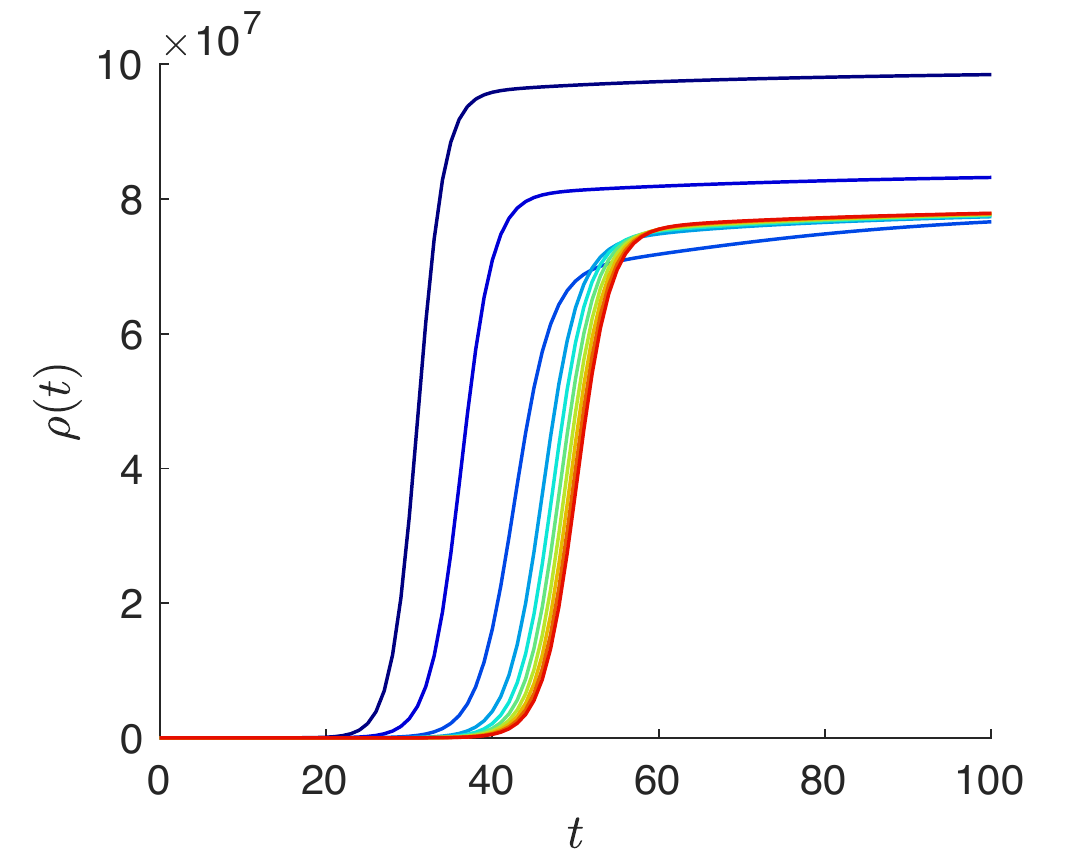} 
    \includegraphics[width=1.0cm]{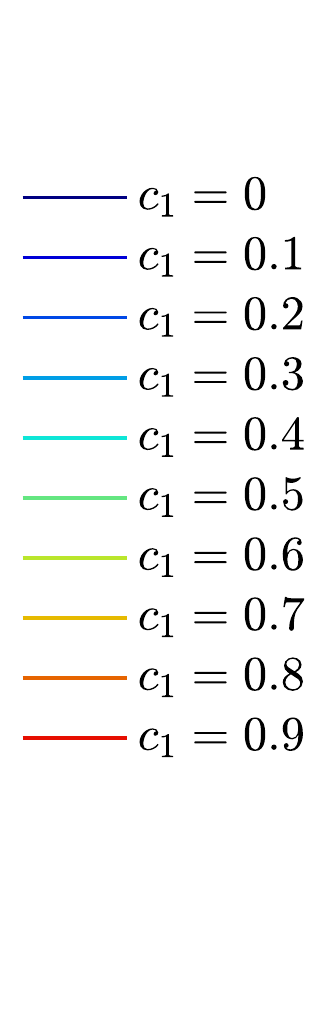} 
    } 
   \caption{ Total number of cancer cells $\rho(t)$ up to $t=100$ 
   simulated with the binary model \eqref{eq:GvrnDiscrete} 
   and continuous model \eqref{eq:Gvrn2}. 
   As the cytotoxic drug is increased, 
   $t_{\rho}^*$ 
   is delayed. 
   The total number of cells at $t_{s}$ when the tumor growth slows down 
   is monotonically reduced as the drug dosage increases 
   in the continuum case (3,i), 
   while it is not in the binary model and case (1,iii). 
   In particular, the dynamics is identical in the binary model 
   when the dosage is relatively high as $c_1 > 0.132$. 
    } 
\label{fig:nTotC_allc1}	
\end{figure} 
In addition to the resistance trait density, 
the following quantities of interest are computed. 
We denote  
the time that the tumor size $\rho(t) = 5 \cdot 10^7$ as 
\begin{equation*}
t_{\rho}^* 
\doteq \min \{ t\, | \, \rho(t) \geq 5 \cdot 10^7 \}.
\end{equation*}
In addition, 
the full cell capacity is approximately computed as  
$ \rho(t_{s})$, 
where $t_{s} \doteq \min \{ t \geq t_{\rho}^*\, | \, 
\rho'(t)/ \rho(t_{\rho}^*) \leq 0.01 \}$, the time when tumor growth slows down.

\begin{figure}[!htb] 
    \centerline{ \footnotesize  (a)   \hspace{3.1cm}   (b)   \hspace{1.3cm}  } 
    \centerline{ 
    \includegraphics[width=3.5cm]{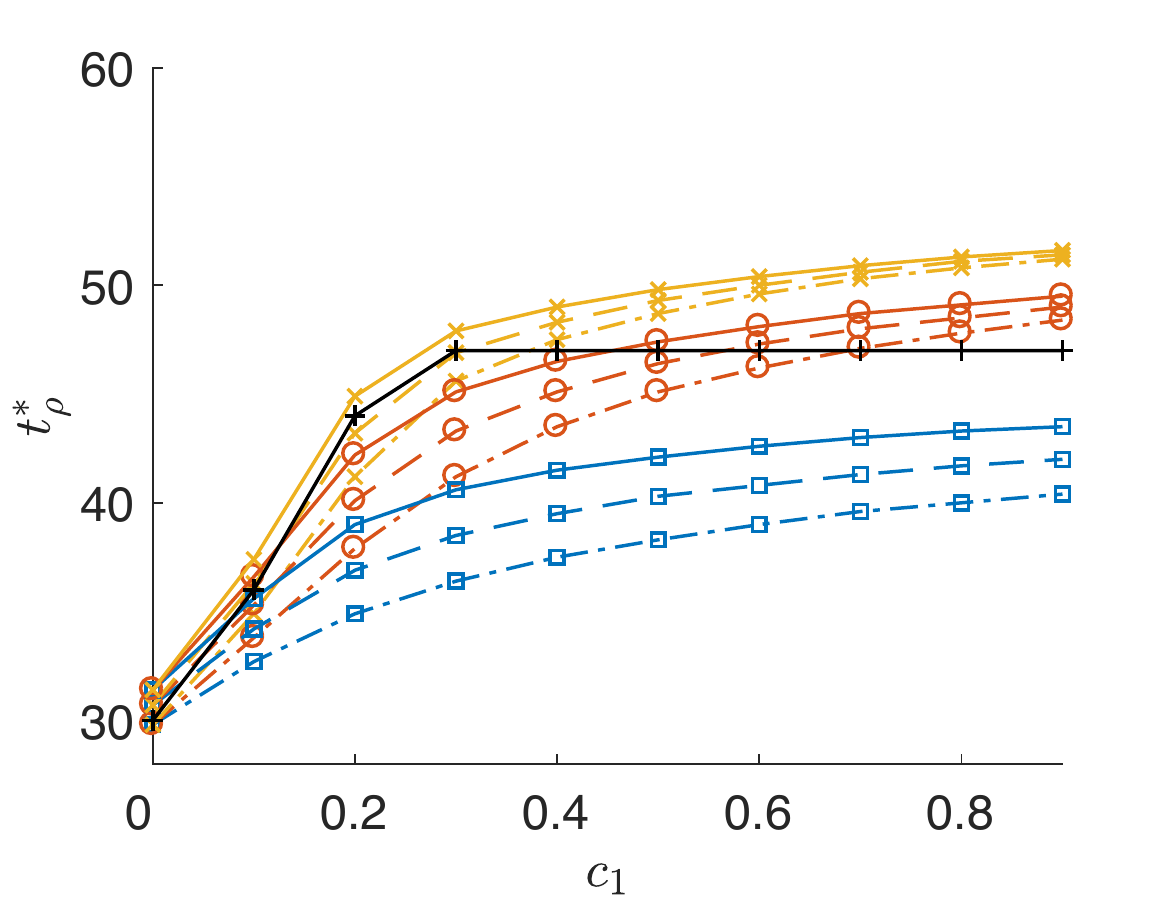}     
    \includegraphics[width=3.5cm]{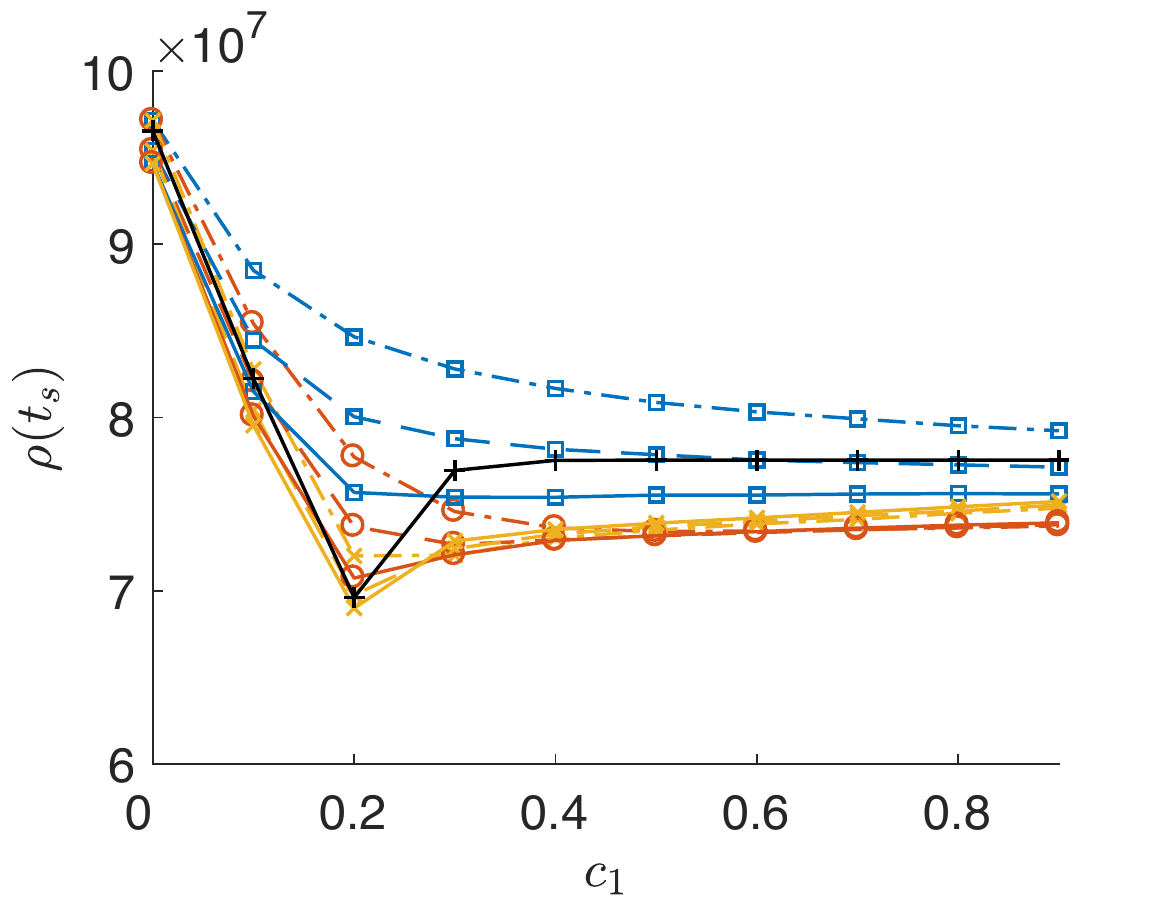} 
    \includegraphics[width=1.2cm]{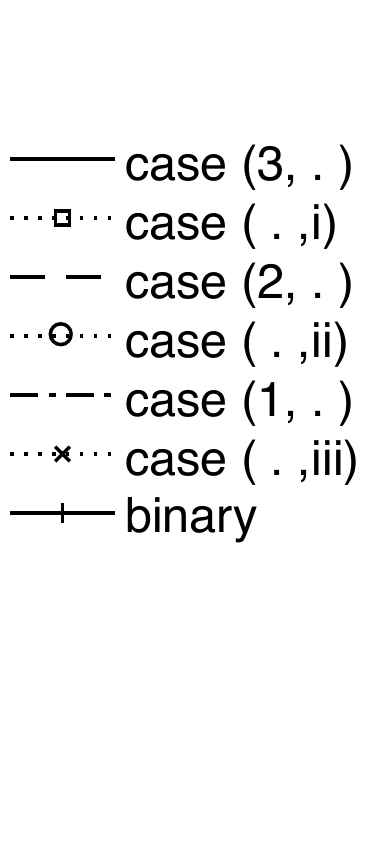} 
    } 
   \caption{ 
   Comparison between the binary model \eqref{eq:GvrnDiscrete} 
   and continuous model \eqref{eq:Gvrn2} 
   regarding the time $t_{\rho}^*$ and cell capacity 
   $\rho(t_{s})$ in terms of cytotoxic drug dosage $c_1$. 
   The binary model yields an identical result 
   when the drug dosage is $c_1 \geq 0.3$, 
   while the results of the continuum models change gradually. 
   Moreover, $t_{\rho}^*$ varies depending on 
   the choice of continuum models 
   and the measured time is shown to be more sensitive 
   to the choice of the drug effect function 
   than to the proliferation function. 
   } 
\label{fig:Time_allc1} 
\end{figure} 
Figure \ref{fig:nTotC_allc1} compares the dynamics of the 
total number of cancer cells $\rho(t)$ using 
the continuous model \eqref{eq:Gvrn2} and binary 
model \eqref{eq:GvrnDiscrete} 
up to $t=100$. 
The times $t_{\rho}^*$ and $t_{s}$ are delayed 
as the cytotoxic drug dosage increases.
However, 
in the binary model, 
the results are essentially identical 
when the dosage is relatively high as $c_1 > \eta = 0.132$. 
Moreover, 
the tumor size of approximate full capacity $\rho(t_{s})$ 
in the continuum case (3,i) 
is gradually reduced as the drug dosage increases, 
which is not the case in the binary-trait model and case (1,iii). 
The results of $t_{\rho}^*$ and  $ \rho(t_{s})$ 
with respect to the cytotoxic drug dosage $c_1$ 
shown in Figure \ref{fig:Time_allc1}, 
where the distinction between the binary and continuum models 
are more apparent. 
The binary model yields an identical result after 
the drug dosage increases above $c_1 \geq 0.3$, 
while the continuum models show a gradual change 
depending on the drug dosage. 
We observe that with our model parameters 
the results are more sensitive to the choice of 
the drug effect function  (case i, ii, iii) 
than to 
the proliferation function (case 1, 2, 3).

\begin{figure}[!thb]
    \centerline{ \footnotesize\rotatebox{90}{\hspace{.7cm} case (3,i)} 
    \includegraphics[width=11cm]{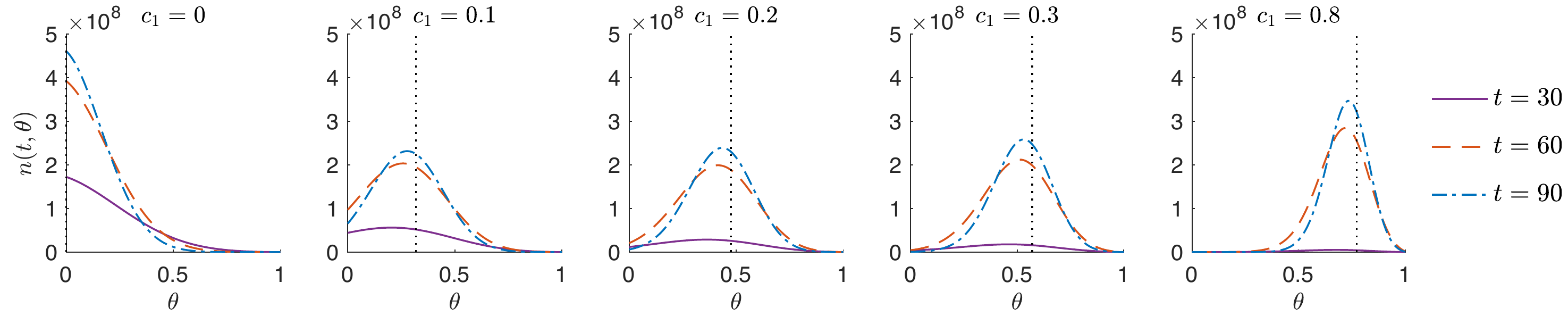} 
    }
    \centerline{ \footnotesize\rotatebox{90}{\hspace{.6cm} case (2,ii)} 
    \includegraphics[width=11cm]{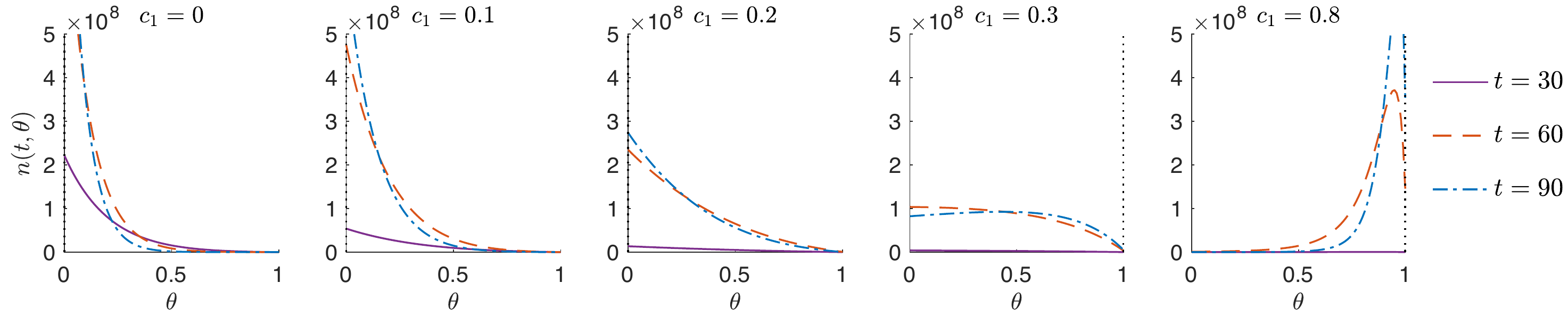} 
    }
    \centerline{ \footnotesize\rotatebox{90}{\hspace{.6cm} case (1,iii)} 
    \includegraphics[width=11cm]{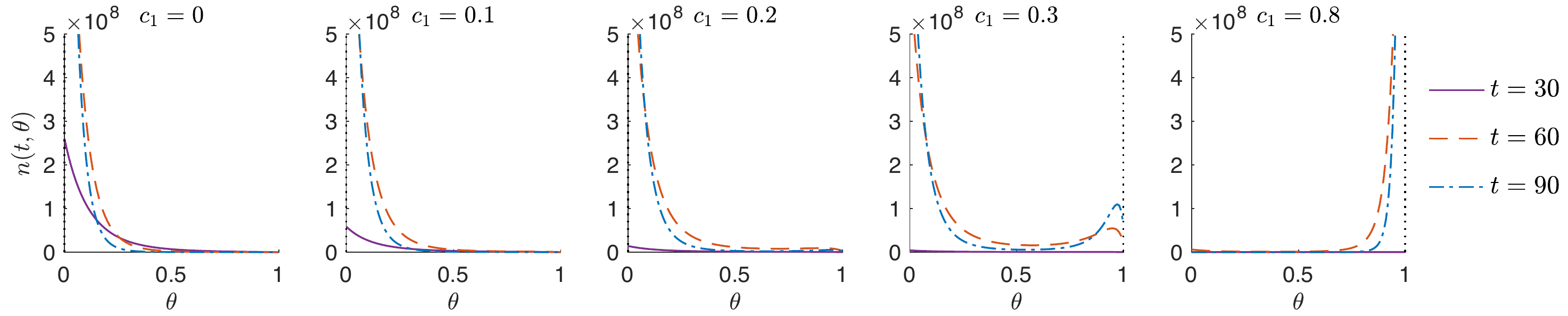} 
    }
   \caption{  The cancer cell distribution using continuum model 
   \eqref{eq:Gvrn2} 
    for different dosages of cytostatic drug 
   at time $t = 30$, 60, 90. The case (3,i) shows a smooth transition of intermediate maximal resistance trait as 
$\theta_M(c_1)  =  3+1/2c_1 - \sqrt{4+3/c_1+1/4c_1^2 }$. 
On the other hand, cases (2,ii) and (1,iii) show 
maximal trait either at the most sensitive or the most resistant trait 
depending on the drug dosage threshold $c_1 = 0.25$ (see Table \ref{Tbl:maxfit2}). 
     } 
\label{fig:Pntp_allc2}	
\end{figure} 
\begin{figure}[!htb] 
    \centerline{ \footnotesize  (a)   \hspace{3.1cm}   (b)   \hspace{1.3cm}  } 
    \centerline{ 
    \includegraphics[width=3.5cm]{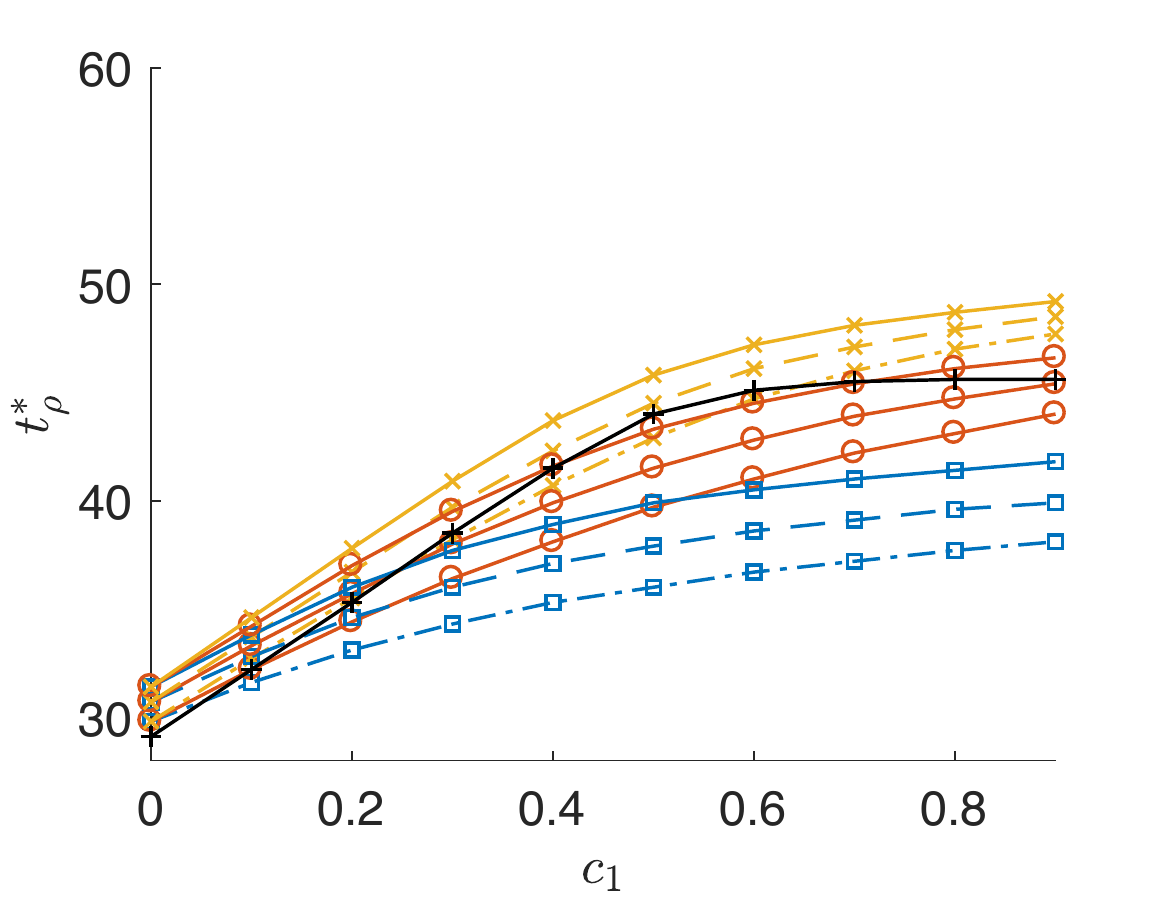}     
    \includegraphics[width=3.5cm]{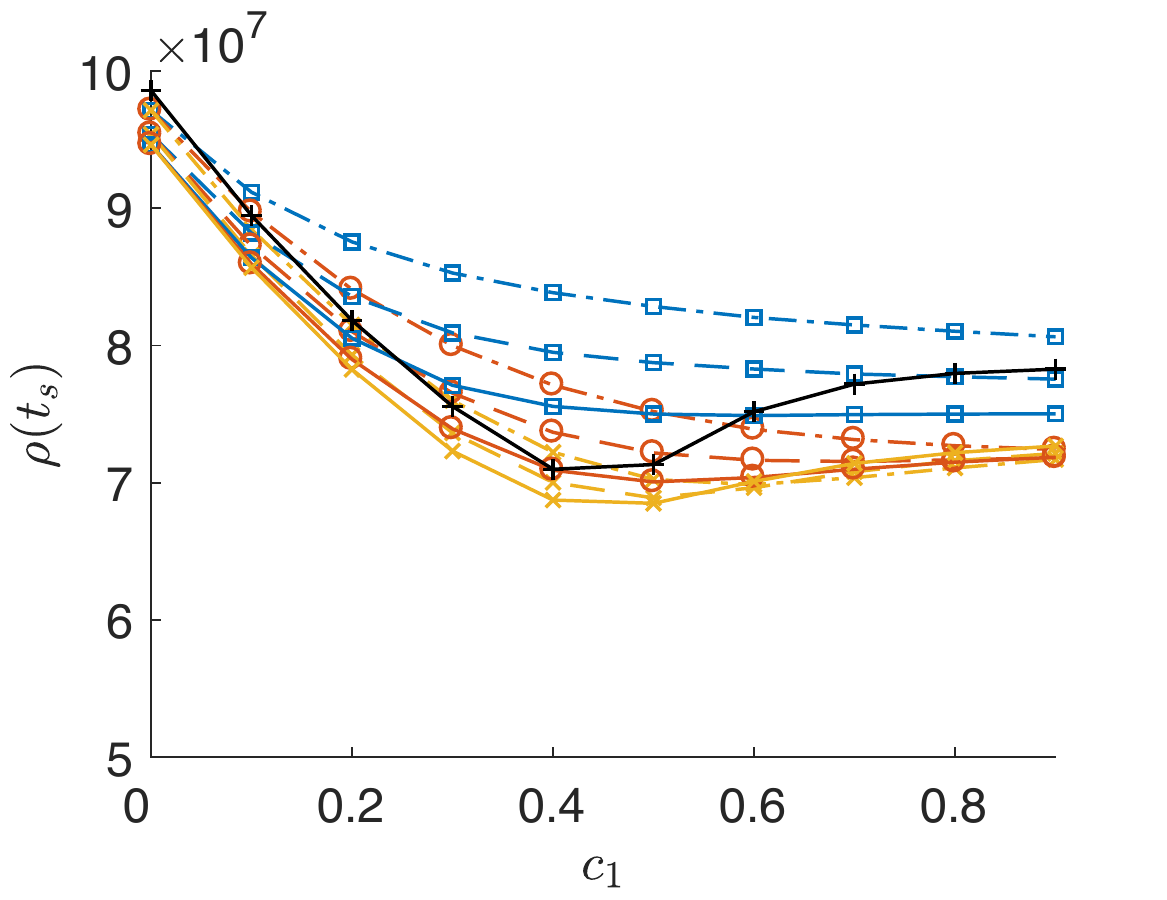} 
    \includegraphics[width=1.2cm]{171226_nTotCvsDrug_legend.pdf} 
    } 
   \caption{ 
   Comparison between the binary model \eqref{eq:GvrnDiscrete_} 
   and continuous model \eqref{eq:Gvrn2} 
   regarding the time $t_{\rho}^*$ and cell capacity 
   $\rho(t_{s})$ with respect to the cytostatic drug dosage $c_1$. 
   In this case, the binary model also yields a gradual change  
   regarding the drug dosage, still it varies from the results of 
   different continuum models. 
   } 
\label{fig:Time_allc2} 
\end{figure} 

The case of a cytostatic drug 
comparing the continuous model \eqref{eq:Gvrn2} 
and binary model \eqref{eq:GvrnDiscrete_} 
is shown in 
Figures~\ref{fig:Pntp_allc2} and~\ref{fig:Time_allc2}.
The resistance trait distribution 
considering cases (3,i), (2,ii), and (1,iii) 
are plotted in Figure \ref{fig:Pntp_allc2}. 
The intermediate resistance level of maximal fitness 
is achieved in case (3,i) for all drug dosages $c_1$ at 
$\theta_M(c_1) =  {C_{1i}}/{2} - \sqrt{{C_{1i}^2}/{4} - 5}$, 
where $C_{1i} = 6+{1}/{c_1}$, 
similar to the results of using cytotoxic drugs. 
We also observe a binary outcome 
either at the most sensitive or the most resistant trait 
depending on the drug dosage threshold 
$c_1 = \eta/(\gamma-\eta) = 0.25$. 
The time $t_{\rho}^*$ and approximate capacity $\rho(t_{s})$ are 
shown in Figure \ref{fig:Time_allc2}. 
In contrast to the cytotoxic drug case,
the binary model also shows a gradual change as a function of the
drug dosage. Still, 
the results obtained by the binary and continuous models are different.

\subsection{Epimutation in drug resistance} \label{sec:2-4}

In this section, we investigate the effect of epimutation on 
the drug resistance dynamics of cancer cells. 
Phenotypic variants in cancer cell populations 
emerge not only from genetic mutations, 
but also due to epimutations. 
Epimutations are heritable changes in gene expression 
that do not alter the DNA, but contribute
to the phenotypic instability 
\cite{Brock2009,Glasspool2006,Gupta2011,Newman2006,Raj2008}. 
Recent experiments demonstrate that such non-genetic instability and phenotypic variability 
allows cancer cells to reversibly transit between different phenotypic states \cite{Chang2006,Pisco2013,Sharma2010} 
and contributes to 
development of resistance to cytotoxic drugs \cite{Chisholm2016,Huang2013}. 
In the continuous phenotypic models, 
epimutation can be readily modeled as 
a diffusion term 
assuming that random epimutations 
yield infinitesimally small phenotypic modifications 
\cite{Becker2011,Navin2014, Lorenzi2016}. 
The dynamics of proliferating cells 
in Eq.~\eqref{eq:Gvrn2} 
with an epimutation rate $\nu$ 
can be written as 
\begin{linenomath}
\begin{equation}
 \partial_t n(t,\theta)  = (R(\theta) - d \rho(t) - C( \theta)) n + \nu \frac{\partial^2 n}{\partial \theta^2 }. 
\label{eq:GvrnDiff}
\end{equation}
\end{linenomath}
The asymptotic distribution of the continuum model 
with epimutation 
for the case (3,i) 
is derived in \citep{Lorenzi2016}.  
Here, we study the effect of epimutation 
in different continuum models. 

\begin{figure}[!htb] 
    \centerline{ \footnotesize \rotatebox{90}{\hspace{0.3cm} case (3,i)} 
    \includegraphics[width=10cm]{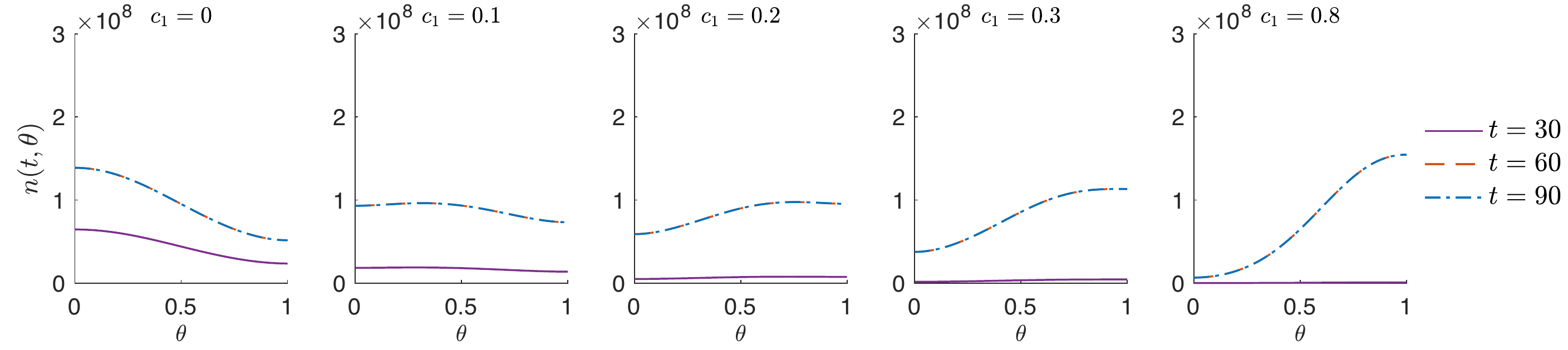} 
    }
    \centerline{ \footnotesize \rotatebox{90}{\hspace{0.3cm} case (1,iii)} 
    \includegraphics[width=10cm]{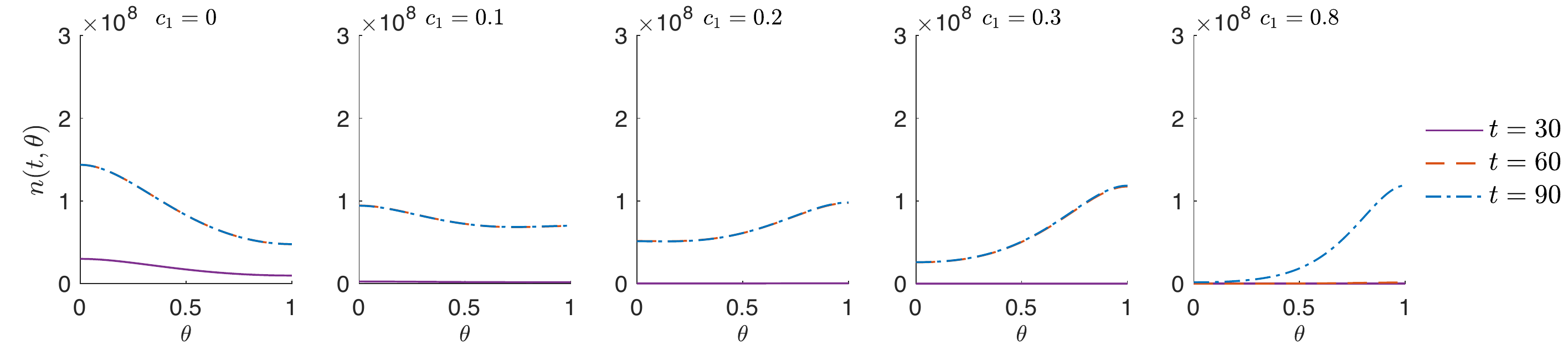}  
     }
   \caption{The cancer cell distribution using continuum model
   \eqref{eq:GvrnDiff} with nonzero epimutation rate $\nu = 10^{-2}$. 
   The results shown are for different drug dosages at times $t = 30$, 60, 90. 
   While the maximal resistant traits are similar to the results without the epimutations as in Figure \ref{fig:Pntp_allc1}, 
   the cell population is significantly more heterogeneous. 
     } 
\label{fig:Conti_epimut}	
\end{figure} 

\begin{figure}[!htb] 
   \centerline{ \footnotesize \hspace{0.2cm} cytotoxic drug \hspace{2.1cm}   cytostatic drug } 
    \centerline{ 
    \includegraphics[width=9cm]{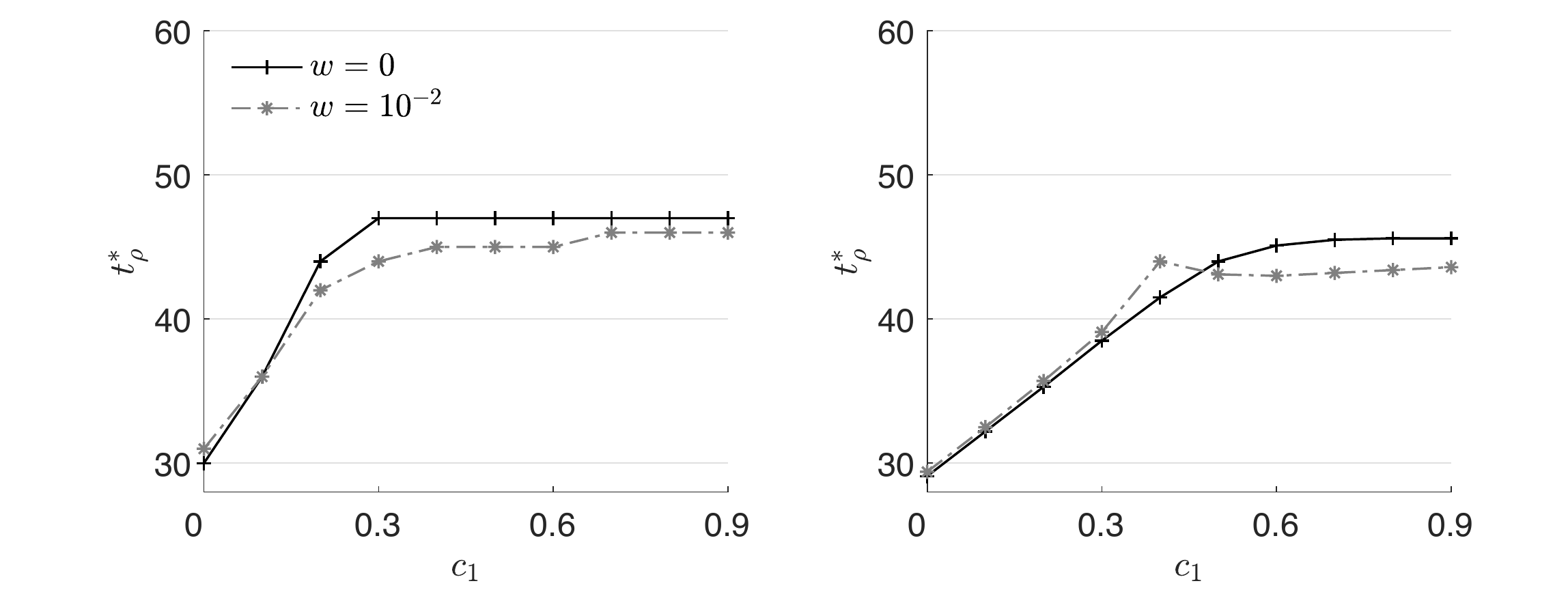} 
    }
   \caption{The comparison of $t_{\rho}^*$ using the binary models \eqref{eq:GvrnDiscrete}--\eqref{eq:GvrnDiscrete_} with 
   mutation rate $w=10^{-2}$ compared with the model with no
   mutations ($w=0$).  In general, mutations 
   result with an earlier relapse due to an increased portion of resistant cells, 
   when the drug dosage is sufficiently high, i.e.,
   $c_1\geq 0.2$ with a cytotoxic drug and $c_2\geq 0.5$ with a cytostatic drug. 
     } 
\label{fig:Dscrt_mut}	
\end{figure} 

Figure \ref{fig:Conti_epimut}	shows the resistance trait 
density $n(t,\theta)$ with epimutation 
using Eq.~\eqref{eq:GvrnDiff} 
corresponding to cases (3,i) and (1,iii) 
when the rate of epimutation is $\nu = 10^{-2}$. 
Although the maximum fitness trait is similar to 
the results without epimutations in 
Figure \ref{fig:Pntp_allc1}, 
the phenotypic instability 
yields a significantly more heterogeneous population, 
not only in case (3,i), where the maximal fitness 
trait is intermediate, 
but also in case (1,iii), where the distribution becomes a Dirac-delta 
function at the boundary trait without epimutations. 

\begin{figure}[!thb]
    \centerline{  \footnotesize  (a)  \hspace{3.5cm}   (b) \hspace{1.6cm} } 
   \centerline{ 
    \includegraphics[width=10cm]{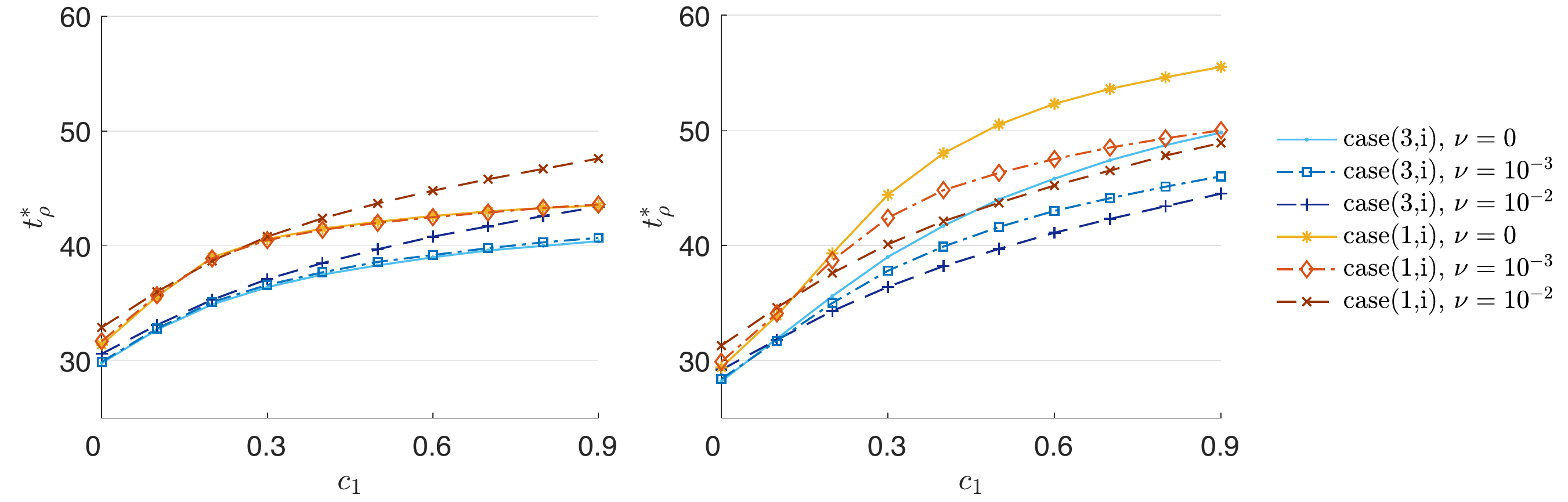} 
    }
   \centerline{ 
    \includegraphics[width=10cm]{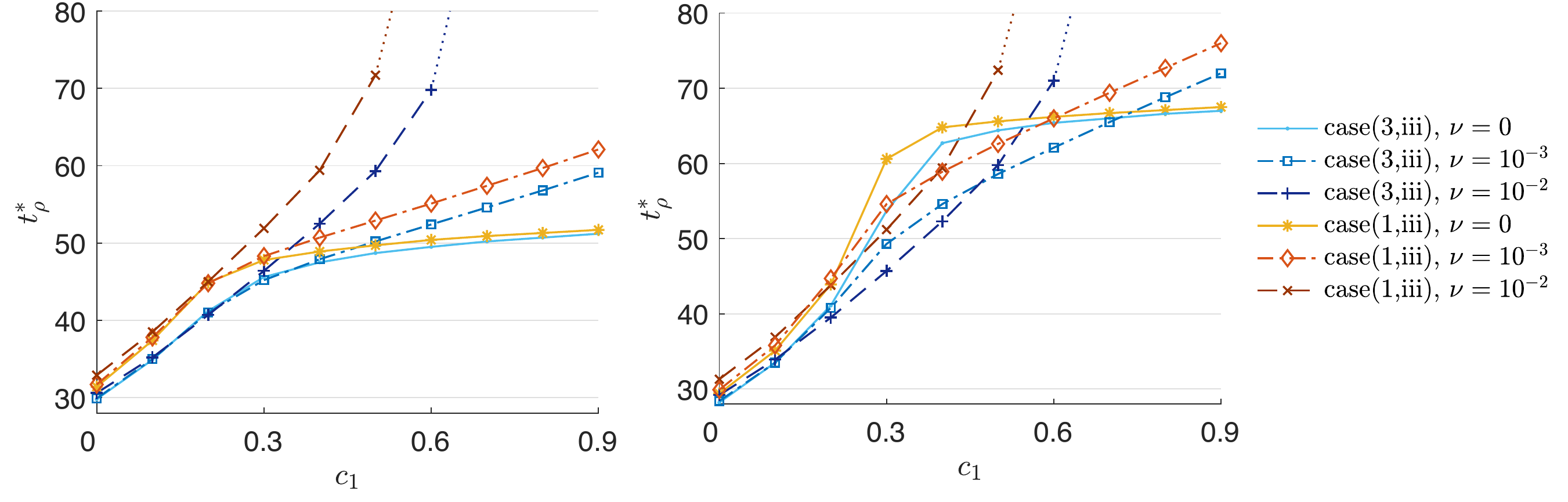} 
    }
   \caption{The comparison of  $t_{\rho}^*$ using the
     epimutation model \eqref{eq:GvrnDiff} subject to cytotoxic drugs. (a) and (b)
     correspond to different amount of preexisting resistance, modeled
     by the initial conditions $n_a(\theta)$ and $n_b(\theta)$, respectively.  
   While mutations in the discrete case accelerate the relapse time, 
   epimutations in the continuum models 
   often delay the relapse time, especially with the initial condition $n_a$. 
   With the initial condition $n_b$, epimutations 
   accelerate the relapse in case (i), 
   but in case (iii) only for a certain range of the drug dosage. } 
\label{fig:Trlsp_epi}	
\end{figure}

\begin{figure}[!thb]
    \centerline{  \footnotesize  (a)  \hspace{3.5cm}   (b) \hspace{1.6cm} } 
    \centerline{ 
    \includegraphics[width=10cm]{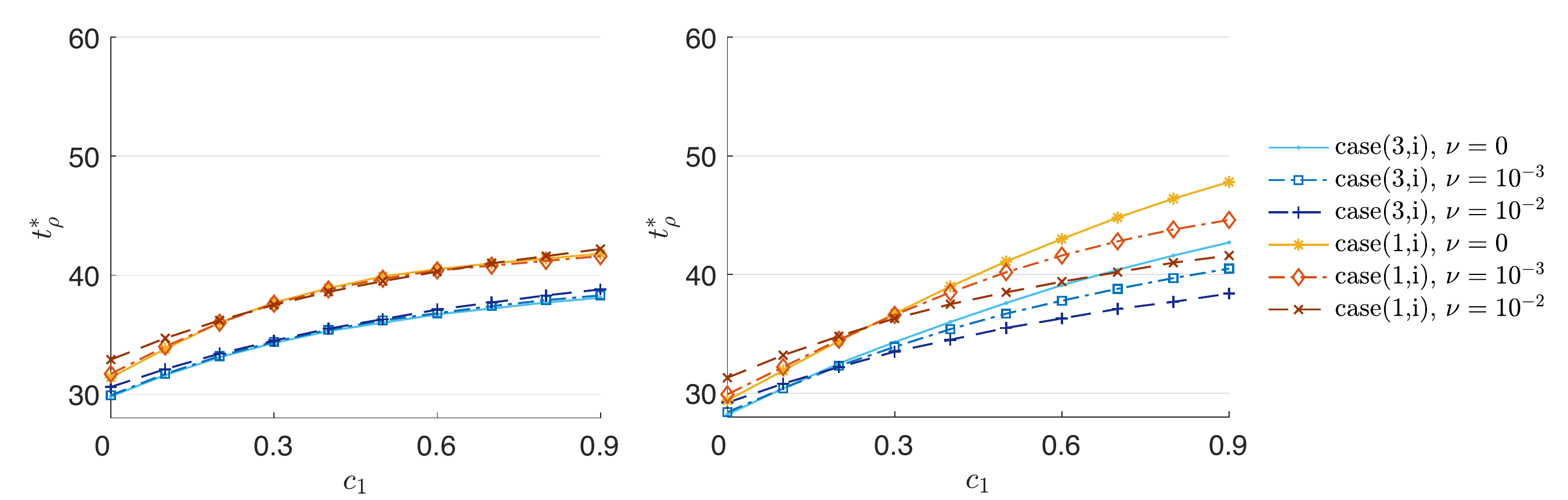}  }
    \centerline{ 
    \includegraphics[width=10cm]{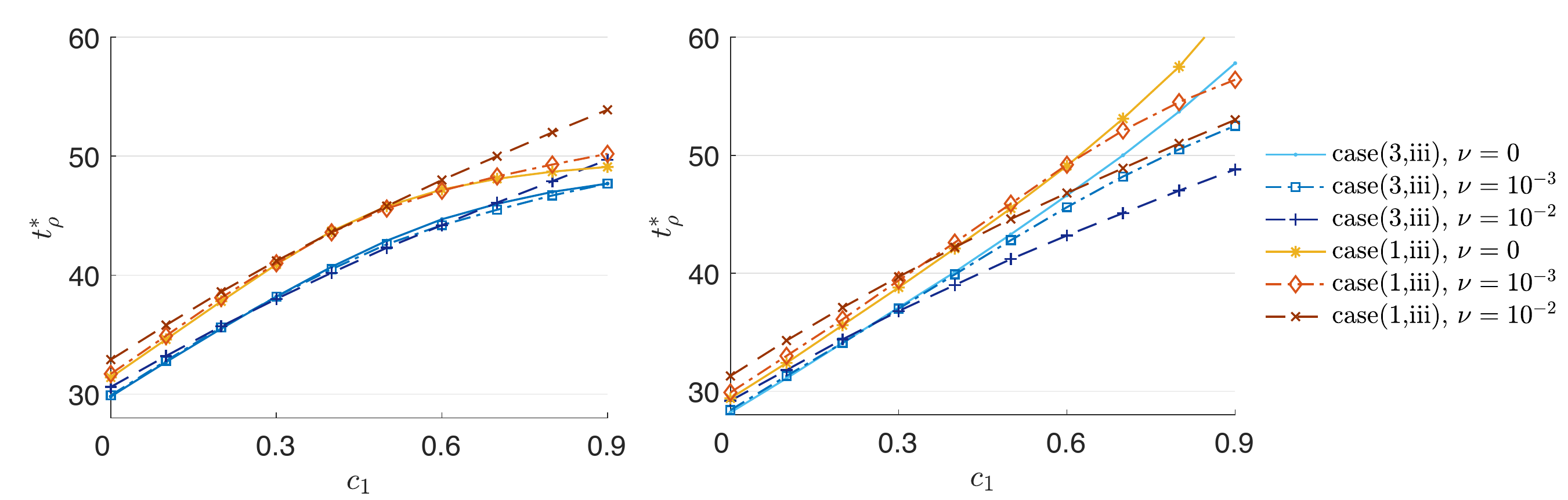}  }  
   \caption{The comparison of $t_{\rho}^*$ using the
     epimutation model \eqref{eq:GvrnDiff} subject to cytostatic drugs. (a) and (b)
     correspond to different amounts of preexisting resistance,
     modeled by the initial conditions $n_a(\theta)$ and $n_b(\theta)$,
     respectively.
    Compared with the cytotoxic drugs, 
    resistance to cytostatic drugs is less affected by 
    epimutations especially with the initial condition $n_a$. 
    However, an earlier relapse is observed with the
    initial condition $n_b$.  
 } 
\label{fig:Trlsp_epi2}	
\end{figure}

We now study the effect of epimutations 
on the time $t_{\rho}^*$ that the tumor size reaches a certain size 
in different models 
subject to cytotoxic drugs. 
In particular, we compare epimutations with regular mutations. 
Figure \ref{fig:Dscrt_mut} shows the time of 
relapse using the binary models \eqref{eq:GvrnDiscrete}--\eqref{eq:GvrnDiscrete_} 
with and without mutations of rate $w=10^{-2}$ initiated from 
$n_S(0) = 0.99$ and $n_R(0) = 0.01$.
In general, mutations accelerate the relapse time
by increasing the proportion of 
resistant cells 
under a sufficiently high dosage. 
We remark that this is similar in the continuum models, 
when using the asymmetric mutation kernel $M(\theta,\vartheta)$ 
described section~\ref{sec:meth}. 
However, Figures \ref{fig:Trlsp_epi} and \ref{fig:Trlsp_epi2} show that 
epimutations in the continuum model \eqref{eq:GvrnDiff} 
often delay the relapse time. 
We consider two initial conditions:
(a) $n_a(\theta) \doteq n_0 \exp\left[ -\theta^2 / l_0 \right]$, 
where we set $l_0 = 0.0739$ and $n_0$ so that
$\int_{0.5}^{1} n(t=0,\theta) d\theta = 0.01$ 
and $\rho(0) = 1$; 
and (b) a linear distribution 
$n_b(\theta) \doteq -0.98 \theta +0.99$, which has a larger population
of resistant cells. 

In Figure~\ref{fig:Trlsp_epi}, 
using the epimutation model \eqref{eq:GvrnDiff} 
subject to cytotoxic drugs, 
we observe that $t_{\rho}^*$ is delayed 
with the initial condition $n_a$, 
especially in case (iii) with a larger rate $\nu$. 
However, epimutations with initial condition $n_b$ 
accelerate the relapse  in case (i), 
and also for a certain range of drug dosages in case (iii). 
For a higher cytotoxic dosage $c_1$ in case (iii), 
the relapse time is again delayed. 
Similarly, Figure~\ref{fig:Trlsp_epi2} shows the effect of epimutations 
on the conitnuum model \eqref{eq:GvrnDiff} 
subject to cytostatic drugs. 
Compared with the cytotoxic drugs, 
resistance to cytostatic drugs is less affected by 
epimutation especially when starting with the initial condition $n_a$. 
However, an earlier relapse is observed with the
initial condition $n_b$ in both models (i) and (iii). 

In conclusion, 
compared with regular mutations that 
give advantage to tumor growth under drug administration, 
epimutations have more diverse effects that can either promote or slow down tumor growth depending on other circumstances, including 
the drug uptake function and the initial conditions.

\section{Simulating tumor growth under multidrug therapy
} \label{sec:Num}

In this section 
we demonstrate how our continuous phenotype
structured modeling framework can be used to 
study MDR. The impact of the tumor's turnover rate and the
proliferating fraction of cancer cells have been 
studied within a discrete phenotype framework  
by Komarova and Wodarz (2005) \citep{komarova2005} and by Gardner (2002) \citep{gardner2002}. 
Here, we compare the results obtained with our approach with 
the conclusions of \citep{komarova2005,gardner2002}.

\subsection{Multidrug resistance: tumor turnover rate} \label{sec:Kmrv}

The impact of the turnover rate in 
tumor growth and resistance dynamics 
has been studied by Komarova and Wodarz (2005)
\cite{komarova2005}.
Their model 
assumes two discrete states for $M$ cytotoxic drugs, 
adding to $2^M$ discrete resistance levels.
The model assumes a constant growth rate $R$,
a constant death rate $D$, and is independent of the cell-cycle.
Komarova and Wodarz
conclude that when comparing tumors of identical sizes at detection,
high turnover tumors ($ R \approx D $)  
have a higher probability of treatment 
failure than low turnover tumors 
($ R \ll D $).
Moreover, a combination therapy ($M>1$) is less likely to have an 
advantage over single-drug therapy in tumors with high turnover rates. 
In contrast, in the continuum models 
we show that depending on the proliferation and drug response functions,
a combination therapy to high turnover tumor 
can be more effective than a single drug treatment.
This is the case with relatively higher dosages when the
drug uptake follows model (i). 
In addition, increasing the dosage in low turnover tumors 
is effective in delaying the tumor relapse 
when the drug uptake follows model (i), 
but not in model (iii).

\begin{figure}[!htb]
   \centerline{ \footnotesize  \hspace{.4cm}  $c = 0.1$ \hspace{1.2cm} $c = 0.2$ \hspace{1.2cm} $c = 0.4$  }
    \centerline{ \footnotesize \rotatebox{90}{\hspace{0.75cm} (iii,iii) \hspace{1.1cm} (i,iii) \hspace{1.25cm} (i,i)  }
    \includegraphics[width=8cm]{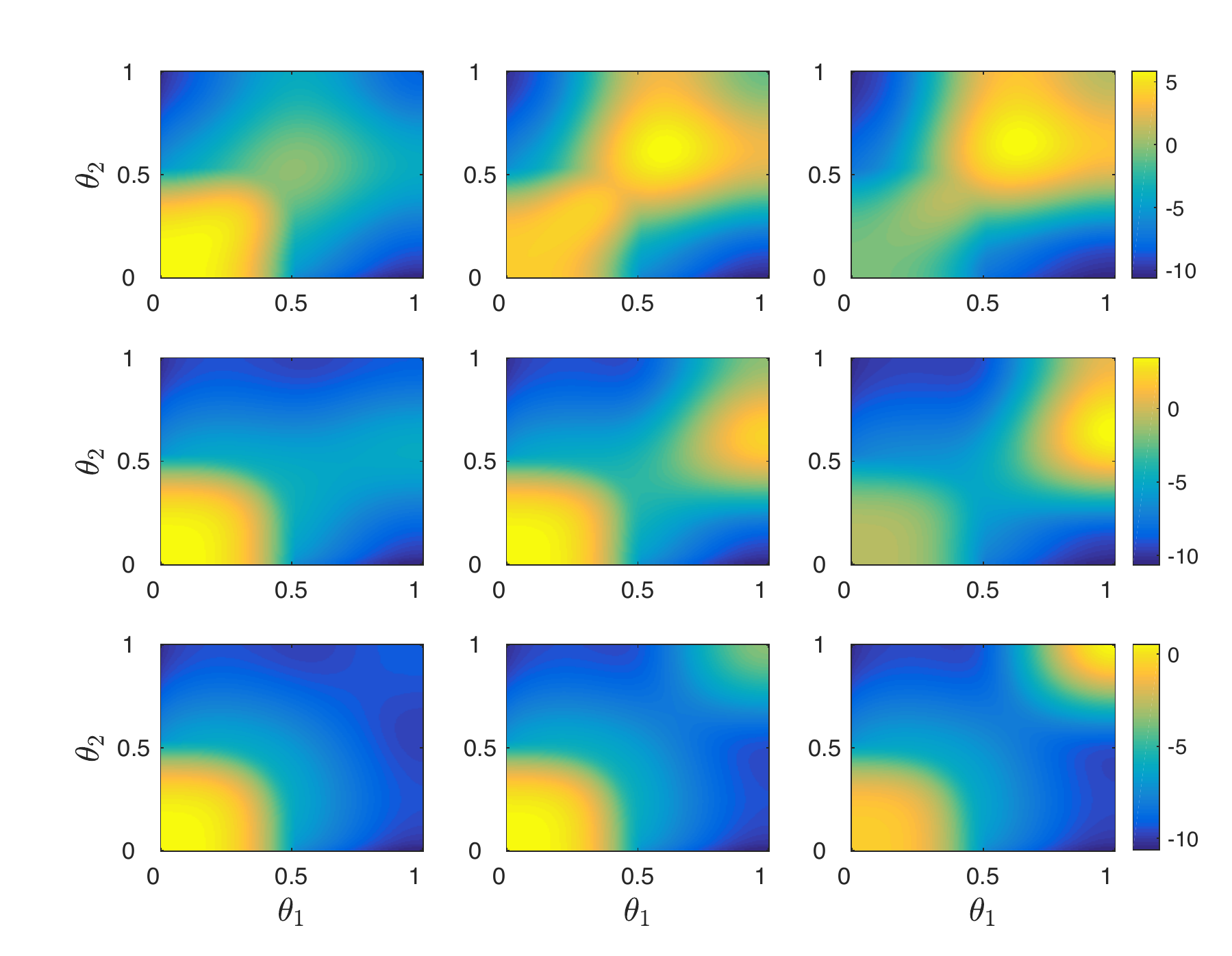} 
    } 
   \caption{Phenotype distribution in the continuum resistant space using 
   two drugs with the drug uptake functions of cases (i,i), (i,iii), and (iii,iii) 
   computed using Eq.~\eqref{eq:Gvrn2}. 
   The distributions shown are cancer cell densities in log scale, $\log( n(t,\theta_1,\theta_2) )$, 
   at time $t=100$ for
   drug dosages $c = 0.1$, 0.2, and 0.4. 
   The distribution is more localized near $\theta_i = 0$ or 
   $1$ in case (iii) compared with case (i). 
   } 
\label{fig:Kmrv_pntp2} 
\end{figure}

The simulation we present is computed 
using the continuum model \eqref{eq:Gvrn2} with the
different  
drug response functions in Table~\ref{Tbl:caseRnC}. 
As in \cite{komarova2005}, 
we assume a constant proliferation rate $R = 1$, 
and model the high and low turnover tumor 
by setting $D = 0.9$ and $D=0.1$, respectively. 
The cytotoxic drug effect is 
taken as $C_P(\theta) = c(t) \Phi(\theta)$, 
where we consider a single parameter $c$ 
for the drug dosage, and 
$\Phi(\theta) = 1 - \prod_{i=1}^M ( 1 - \mu_i(\theta)) $ 
with the uptake functions $\mu_i(\theta)$.
We consider the drug dosages around  
$c(t) \approx 0.1 $ in high turnover tumors and 
$c(t) \approx 0.9 $ in low turnover tumors.

Figure \ref{fig:Kmrv_pntp2} presents 
the cell density in the resistance trait space 
using the continuum model \eqref{eq:Gvrn2}  subject to a  
combination therapy using two cytotoxic drugs ($M=2$). 
We consider a high turnover tumor 
with the uptake functions 
of cases (i,i), (i,iii), and (iii,iii), 
and set  the 
drug dosage as $c = 0.1, 0.2, 0.4$. 
The distributions shown are cancer cell densities in log scale, 
$\log( n(t,\theta_1,\theta_2) )$, at time $t=100$.  
The marginalized distribution in each resistance trait 
is similar to the results of section~\ref{sec:2-3}, 
where case (iii) yields 
more localized distributions near $\theta = 1$ 
in relatively higher dosages 
compared to case (i).

\begin{figure}[!htb]
    \centerline{ \rotatebox{90}{\footnotesize \hspace{0.7cm} $D=0.9$ \hspace{1.4cm} $D=0.1$ }    
    \includegraphics[width=11cm]{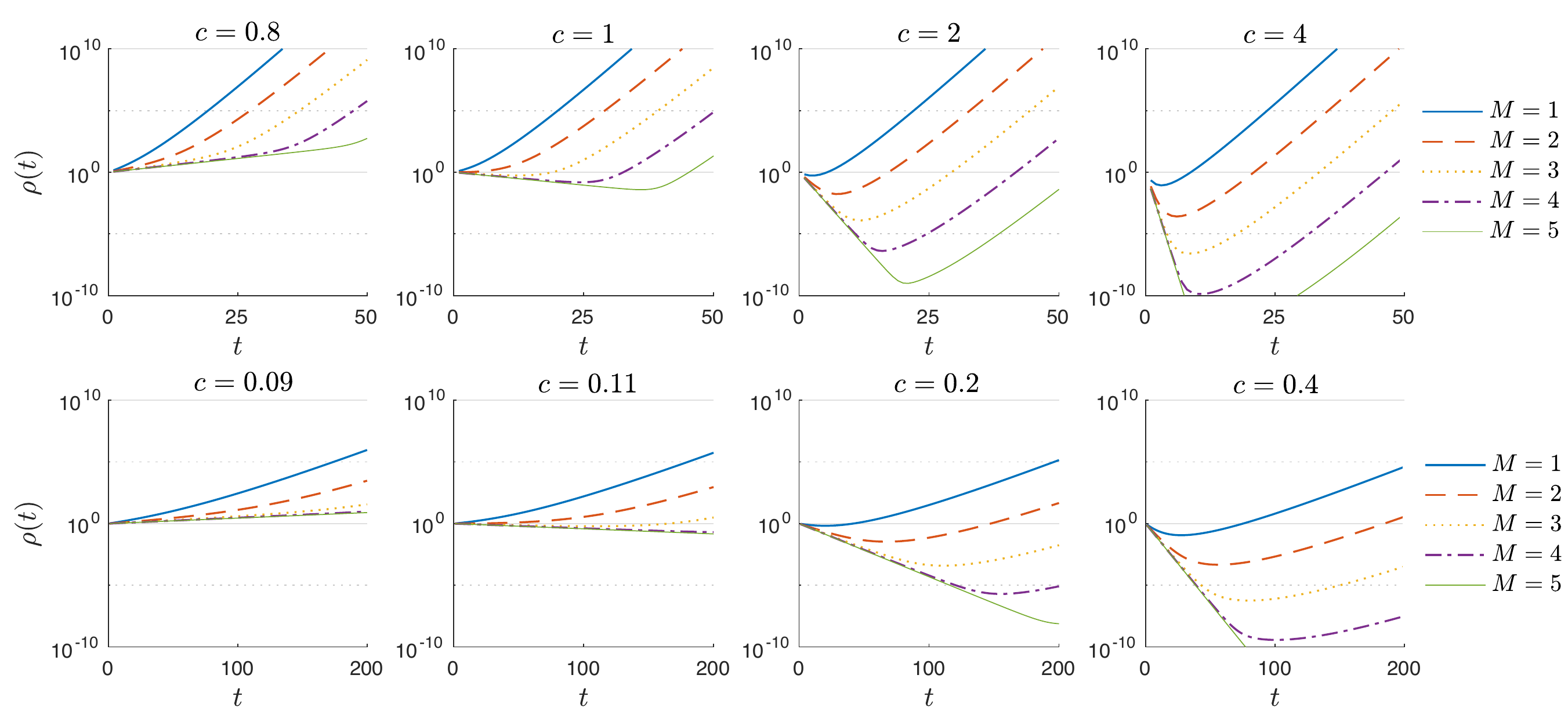}} 
   \caption{The total number of cancer cells $\rho(t)$ using the continuum 
   model \eqref{eq:Gvrn2} and case (i) with $M=1,...,5$ cytotoxic drugs. 
   As the drug dosage $c$ and the number of drugs $M$ are increased, 
   the relapse time is delayed. 
   Increasing the number of drugs to $M \geq 2$ is effective 
   not only in low turnover rates but also 
   in the high turnover rates with relatively high dosages 
   $c \geq 0.2$.
   } 
\label{fig:Kmrv_allM} 
\end{figure}

\begin{figure}[!htb]
    \centerline{ \footnotesize \rotatebox{90}{\hspace{0.9cm} $M=3$ \hspace{1.4cm} $M=1$ }    
    \includegraphics[width=10cm]{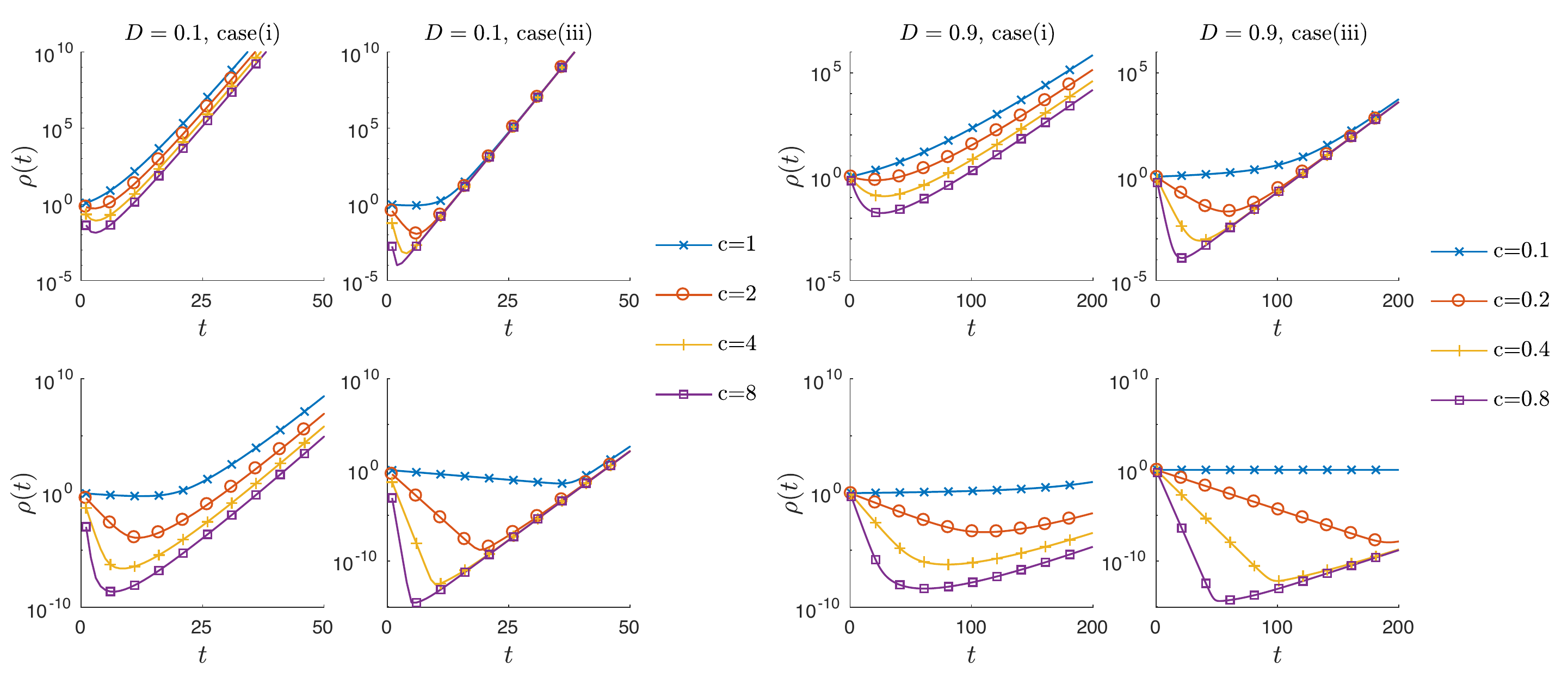} 
    \includegraphics[width=1.7cm]{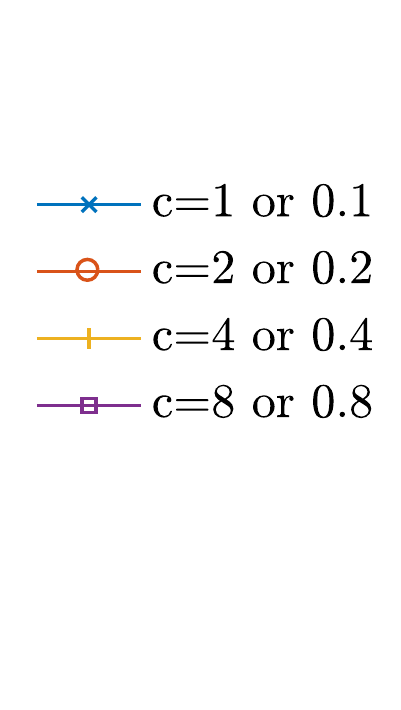}}     
   \caption{Comparison of the number of cancer cells $\rho(t)$ 
   using the continuum model \eqref{eq:Gvrn2} 
   while increasing the drug dosages for different
   turnover rates and drug uptake response models.
   Increasing the dosage is effective when 
   the drug uptake follows model (i), but not in model (iii) regarding the tumor relapse, 
   particularly  
   in low turnover tumors ($D=0.1$). } 
\label{fig:Kmrv_allmodel}
\end{figure}

We now compare the responses of 
high and low turnover tumors with respect to 
the number of drugs $M$ in the continuous models. 
Figure \ref{fig:Kmrv_allM} shows the total number of cells 
$\rho(t)$ up to $t=100$ 
for an increasing 
number of drugs $M=1,\ldots,5$, 
and increasing drug dosages. 
We choose case (i) for the drug uptake function. 
As expected, we observe a delayed growth with an increased
number of drugs and increased dosages.
While increasing the number of drugs  
is not effective in high turnover tumors 
in the model of \cite{komarova2005},  
it is effective 
in the continuum model \eqref{eq:Gvrn2} with the drug update model (i) 
and high dosages $c \geq 0.2$. 
Figure \ref{fig:Kmrv_allmodel} compares 
the total number of cells  
in four different continuum models, 
combining the drug effect (case (i), (iii)) 
and the turnover rate ($D=0.9$, $0.1$). 
We observe that increasing the drug dosage 
over a certain threshold is less likely to 
delay the relapse time 
in low turnover tumor for which the drug uptake follows case (iii).
It is effective in drug uptake case (i).

\begin{figure}[!htb]
    \centerline{ \rotatebox{90}{ \footnotesize \hspace{0.7cm} case (iii) \hspace{1.4cm} case (i) }  
    \includegraphics[width=9cm]{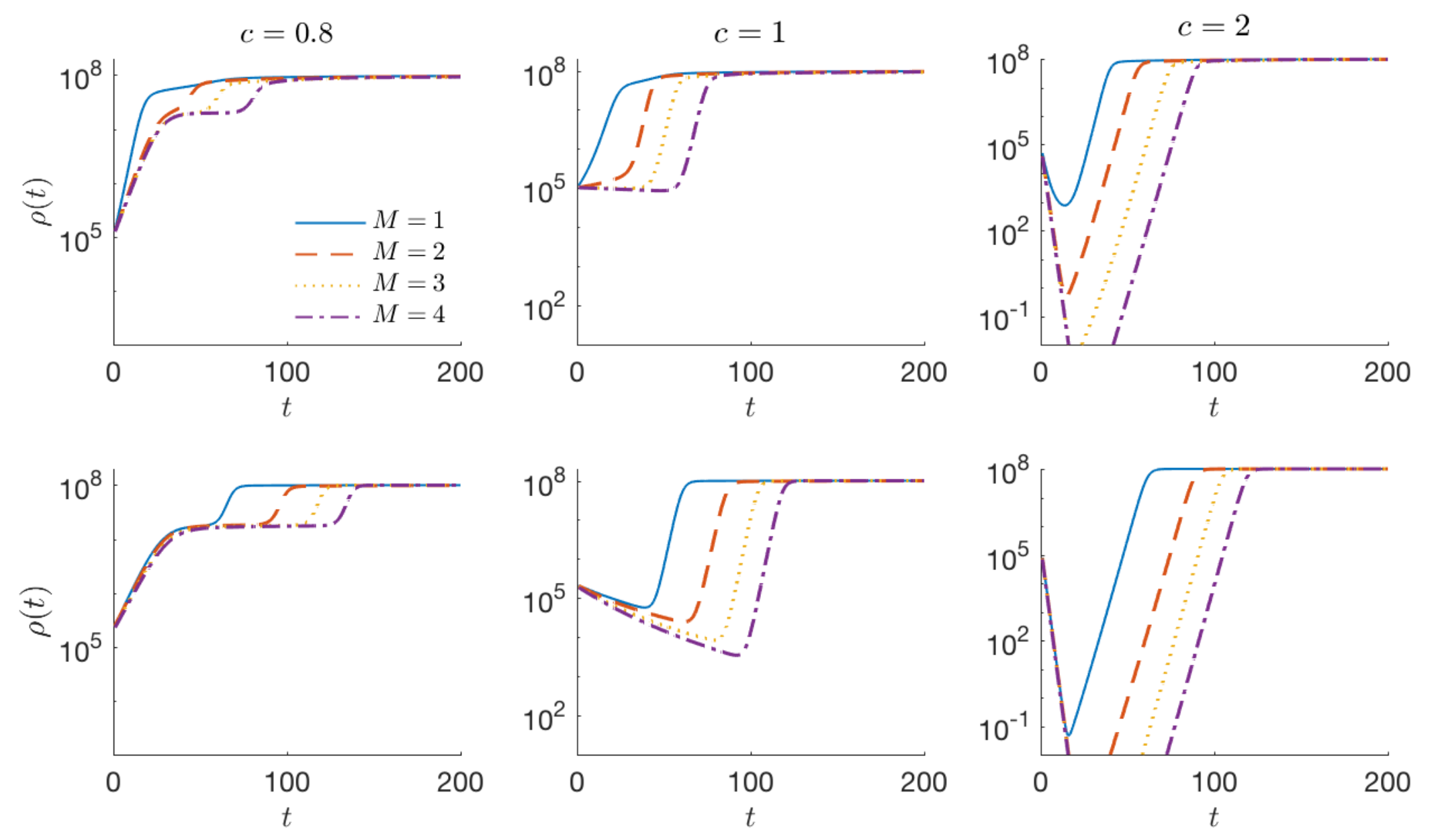}} 
   \caption{ Comparison of the number of cancer cells $\rho(t)$ 
   for an increasing the number of drugs $M$ in the logistic growth model 
   $D = d \rho(t)$ and the continuum model \eqref{eq:Gvrn2}. 
   Assuming a logistic growth, the relapse does not depend on the turnover rate, but on the choice of continuum uptake models. 
   Increasing the number of drugs is more effective in case (iii)
   compared with case (i) in our model. 
   } 
\label{fig:Kmrv_logG}
\end{figure}

Finally, Figure \ref{fig:Kmrv_logG} 
shows the effect of increasing the number of drugs
assuming a logistic growth model by taking $D = d \rho(t)$ 
in Eq.~\eqref{eq:Gvrn2}. 
In this case, the dynamics does not depend on 
the turnover rate $d$ 
except that the cell capacity changes. 
The results are shown for $d = 10^{-8}$, 
and we remark that taking $d = 9 \cdot 10^{-8}$ 
shows essentially no difference. 
However, the relapse does depend on the 
choice of a continuum model. 
Increasing the number of drugs 
delays the relapse in both cases (i) and (iii), 
but more so in case (iii) compared with (i). 

We conclude that 
in addition to the turnover rate,
the drug uptake function of the continuum model 
is also important in controlling the outcome of the treatment.
In particular, 
a combination therapy with multiple drugs is  
effective not only in low turnover tumors, 
but also in high turnover tumors with the drug uptake case (i). 
Moreover, a high cytotoxic drug dosage 
in low turnover tumor with case (iii) is less effective 
than case (i). 
The drug uptake function is often more important
than the turnover rate  
in determining the outcome of the tumor growth and relapse, 
particularly with a logistic growth condition.

\subsection{Multidrug resistance: heterogeneity due to the proliferating index} 
\label{sec:Gardner}

Gardner (2002)  \cite{gardner2002} proposed an individually tailored model 
based on the tumor cell kinetics of patients 
following heterogeneous colonies of 
proliferating and quiescent cells. 
This study considered multidrug
resistance to six specific drugs, including 
two cell-cycle specific (CS) cytotoxic drugs,
5-Fluorouracil and Methotrexate, 
that only affect the proliferating cells;
two cell-cycle nonspecific (nCS) cytotoxic drugs, Cyclophosphamide and Doxorubicin, 
that kill both proliferating and quiescent cells;
and two cytostatic drugs, Tamoxifen and Herceptin. 
The model assumed discrete levels of resistance 
in addition to the parameters of 
cell division rates, apoptotic rates, response to drugs, 
and evolution of drug resistance. 
It then used the discrete model to predict 
drug combinations and schedules 
that are likely to be effective in reducing the tumor size.

\begin{figure}[!htb]
    \centerline{ 
    \includegraphics[width=10cm]{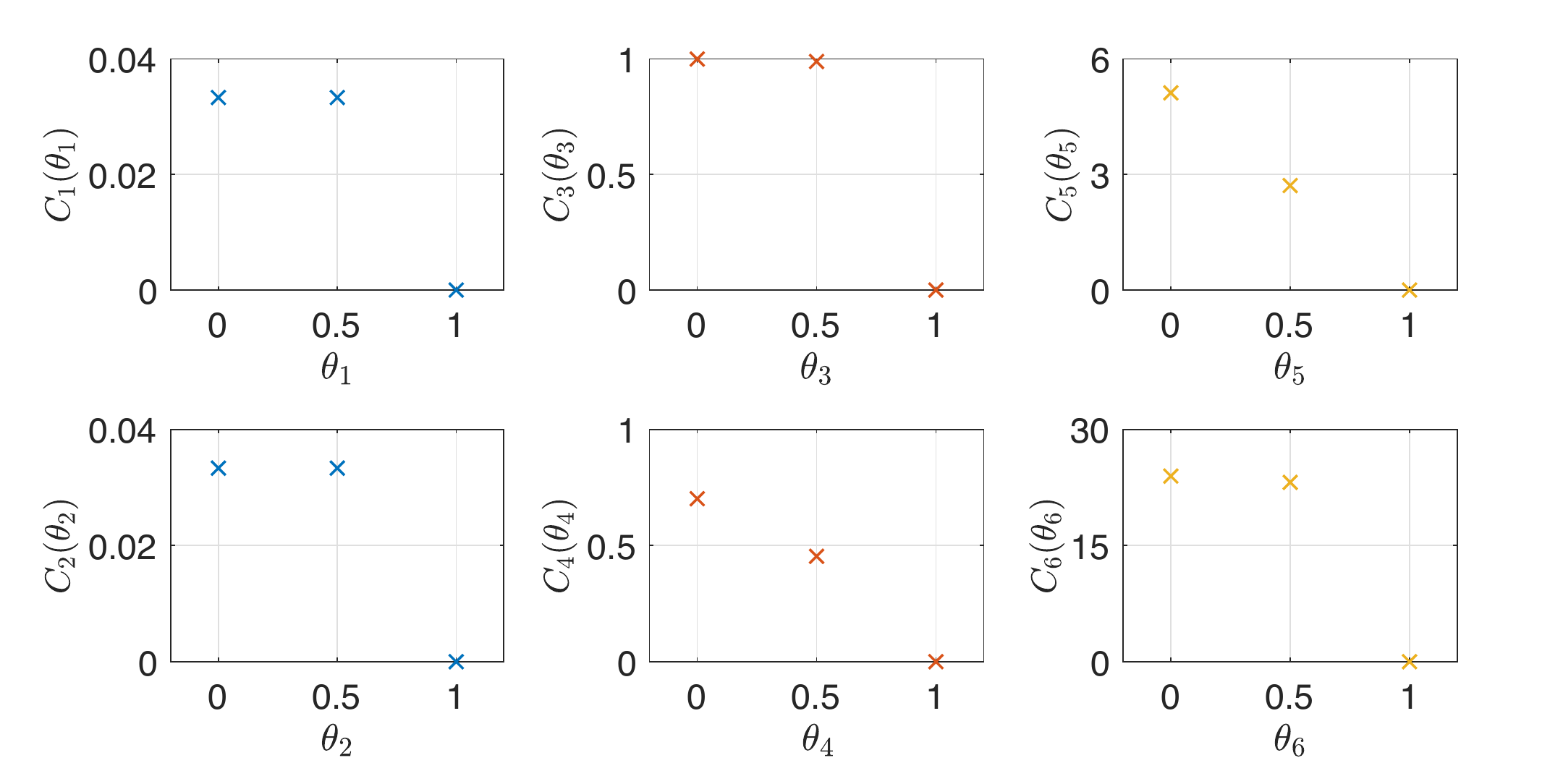} } 
   \caption{ The drug effect $C_i(\theta_i)$ at resistance 
   level $\theta_i \in \{ 0,\, 0.5,\, 1\}$ of the six drugs used in \cite{gardner2002}. 
    The drugs include two CS cytotoxic drugs: 1) 5-Fluorouracil, and
    2) Methotrexate;
    two nCS cytotoxic drugs:  3) Cyclophosphamide, and 4) Doxorubicin; 
    and two cytostatic drugs: 5) Tamoxifen, and 6) Herceptin.
    The exponential kill models can be categorized into the continuum models of cases (ii) and (iii). 
} 
\label{fig:Grdnr_C}
\end{figure}

The governing system in \cite{gardner2002} 
assumes three discrete drug resistance levels, 
$\theta_i = \{ 0,\, 0.5,\, 1\}$,  
for each of the six drugs, 
and it is 
similar to Eqs.~\eqref{eq:Gvrn0}-\eqref{eq:Gvrn1}:
\begin{equation}
\begin{aligned}
	\dot{n}_P &=  \left( (1-w)R - C_P - q
	 \right) n_P + p n_Q + w 
	 \mathcal{M}( n_P),  \\  
	\dot{n}_Q &=  q n_P + \left( -p - D_Q - C_Q 
	 \right) n_Q. 
\end{aligned}  
\label{eq:GvrnGardner}
\end{equation}
Here ${n}_P$ and ${n}_Q$ are defined on $3^6$ discrete resistance levels. 
In addition, $C_P$ includes the effect of apoptosis of proliferating cells 
of rate $D$,  
the quiescent cells die as a result of 
necrosis of rate  $D_Q$, 
and $\mathcal{M}$ denotes the mutation term similar to Eq.~\eqref{eq:Gvrn0} \cite{gardner2002}. 
The transfer rates  from the quiescent cells to the proliferating cells 
to balance a fixed ratio of proliferating cells $\delta^*$ is 
$q = (R-D+D_Q) (1-\delta^*) + {p(1-\delta^*)}/{\delta^*}$. 
We denote the CS cytotoxic drugs as $C_1$ and $C_2$, 
the nCS cytotoxic drugs as $C_3$ and $C_4$, and the cytostatic drugs as $C_5$ and $C_6$. 
The drug effects are modeled using the exponential kill model \cite{Gardner2000a} as
$C_{i}(\theta_i) = R \left[ 1-e^{-a_i (\theta_{max}-\theta_i) c_i(t) } \right] $  
for the CS cytotoxic drug ($i=1,\,2$),  
$C_{i}(\theta_i) = 1-e^{-a_i (\theta_{max}-\theta_i) c_i(t) } $ 
for the nCS cytotoxic drug ($i=3,\,4$), 
and 
$C_{i}(\theta_i) = z_{i} \left[ 1 - e^{-a_i (\theta_{max}-\theta_i) c_i(t) } \right] $ for the cytostatic drug ($i=5,\,6$), 
where $\theta_{max} = 1$ and 
the domain of resistance trait is taken at three discrete levels 
$\theta_i \in \{ 0,\,0.5,\,1\}$. 
The net drug effects are taken as 
\begin{equation}
\begin{aligned}
   C_P(t,\theta) = 1 - (1-D) \prod_{i=1}^4 \left( 1-C_i(\theta_i;c_i(t)) \right), \hspace{2cm} \\ 
  C_Q(t,\theta) = 1 - \prod_{i=3}^4 \left( 1-C_i(\theta_i;c_i(t)) \right), \qquad 
    R(t,\theta) = \dfrac{\varphi(\theta)}{ 1 + \sum_{i=5}^6 C_i(\theta_i;c_i(t))  }. 
\end{aligned} \label{eq:netGardner}
\end{equation}
Figure \ref{fig:Grdnr_C} shows the three 
discrete levels of drug effect 
using the dosages $c_1$, ..., $c_6$ from \cite{gardner2002} (see \ref{sec:appxA}). 
We note that although Gardner considers 
three levels of resistance, the cells with sensitive levels 
$\theta_i=0$ and $\theta_i=0.5$ 
of $i = 1,\,2,\,3$, and $6$ have similar response to the drug. 
Moreover, the exponential kill model of 
$C_1$, $C_2$, $C_3$, and $C_6$ based on the concavity can be classified 
as our case (iii), and 
$C_4$ and $C_5$ as case (ii). 
In the following simulations, 
we assume that the proliferation $R$ and the drug effects $C_i$ 
in Eqs.~\eqref{eq:Gvrn0}--\eqref{eq:Gvrn1} 
follow the models as in Table \ref{Tbl:caseRnC} with the net drug effect as in \eqref{eq:netGardner}, 
and compare the results with the discrete model \eqref{eq:GvrnGardner}. 
See \ref{sec:appxA} for the model parameters.

\begin{figure}[!htb] 
    \centerline{  \footnotesize \hspace{0.7cm} no drug \hspace{2.4cm} $[C_2]$  \hspace{2.1cm} $[C_1,...,C_6]$ } 
    \centerline{ 
    \includegraphics[width=10.2cm]{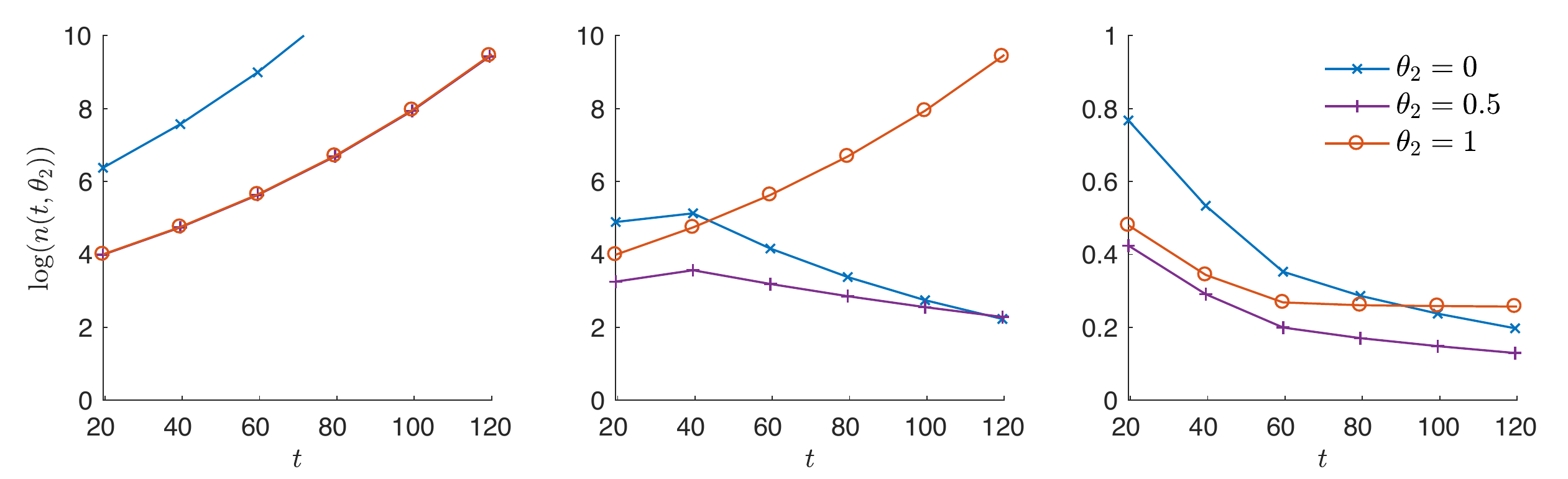} }
    \centerline{ 
    \includegraphics[width=10.2cm]{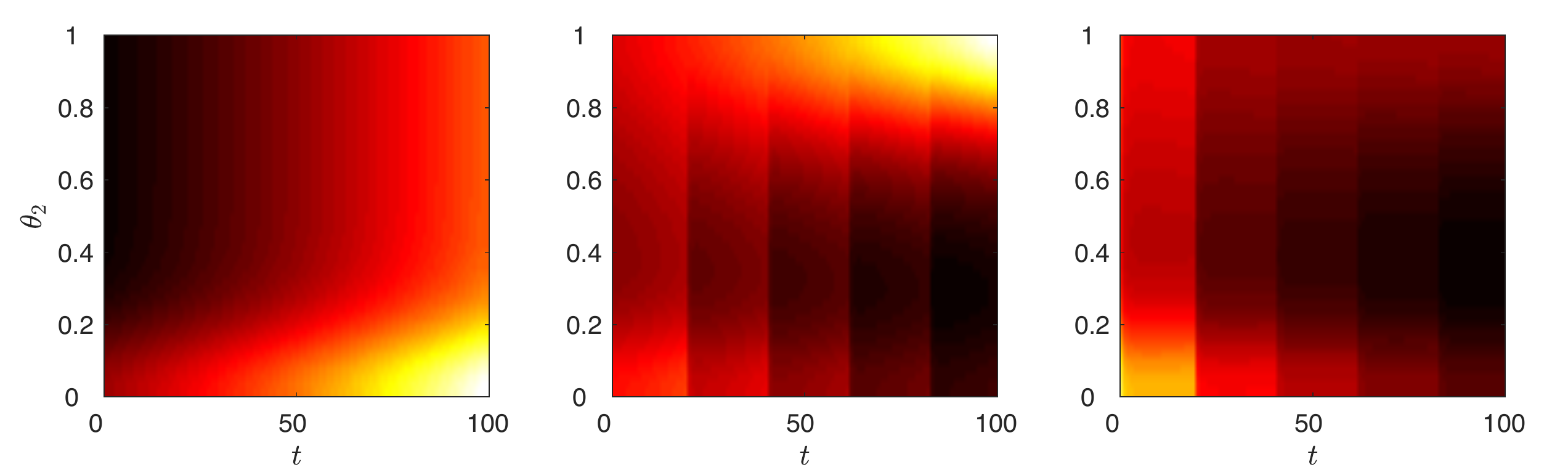} }
   \caption{ 
The cell distribution in the resistance trait space of $C_2$ 
   in log scale, $\log(n(t,\theta_2))$, using no drug, 
   a single drug of $C_2$, and all drugs. The plots compare 
   the discrete model \eqref{eq:GvrnGardner} (top) and the continuum
   model \eqref{eq:Gvrn0}--\eqref{eq:Gvrn1} (bottom). Due to the shape
   of the exponential kill model (case (iii)), the cell distribution
   of the continuum model is concentrated
   at the boundary traits similarly to the discrete model. However, 
   the continuum model reveals the 
   cell distribution in the intermediate levels and 
   the degree of heterogeneity in the resistance trait. 
    } 
\label{fig:Grdnr_Pntp}
\end{figure}

\begin{figure}[!htb]
    \centerline{  \footnotesize  \hspace{0.6cm} [$C_1$, $C_3$] \hspace{1.8cm} [$C_3$, $C_6$] \hspace{1.9cm} [$C_1$,...,$C_6$] \hspace{1.7cm} } 
    \centerline{  
    \includegraphics[width=6cm]{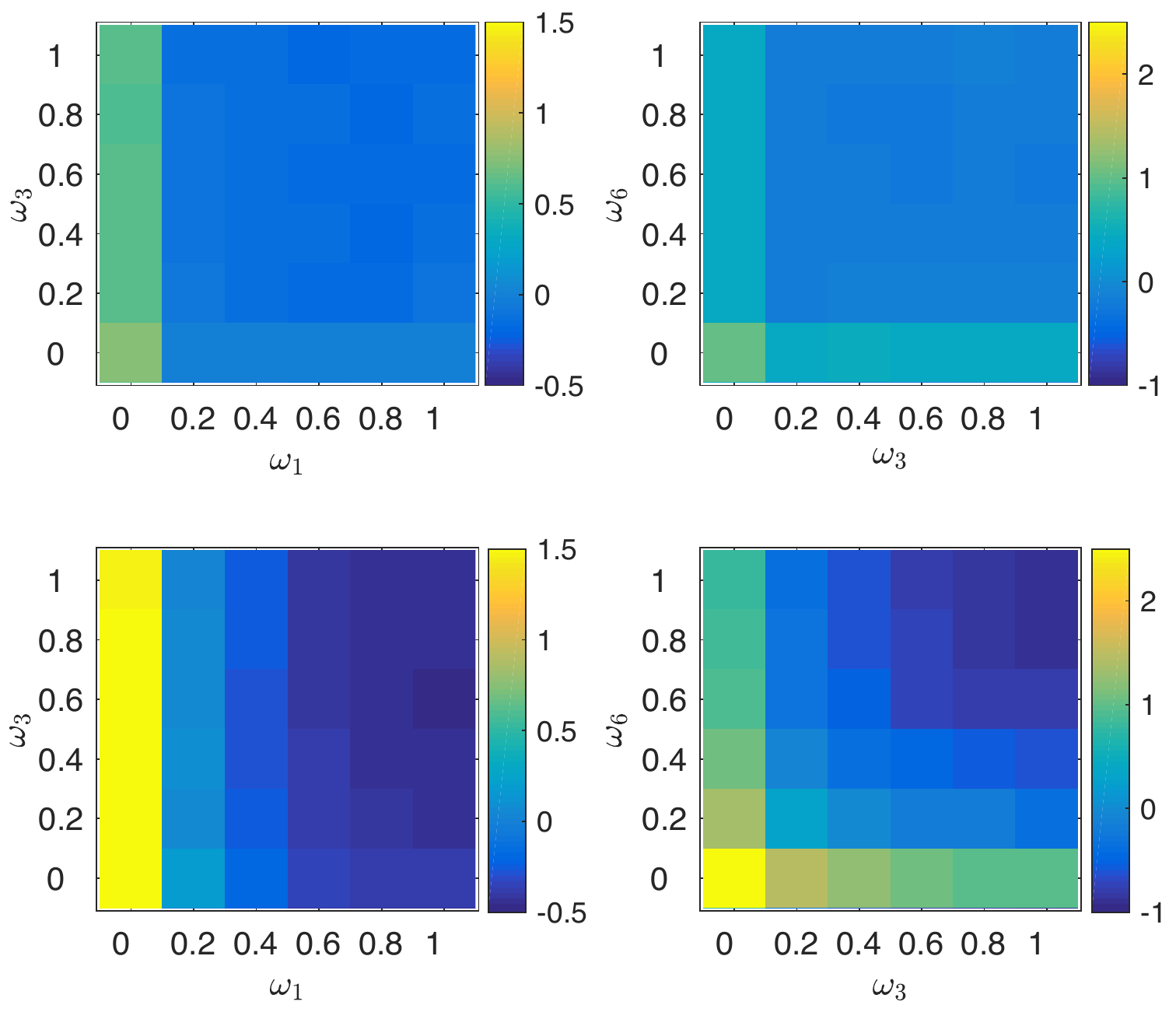}  
    \includegraphics[width=4.5cm]{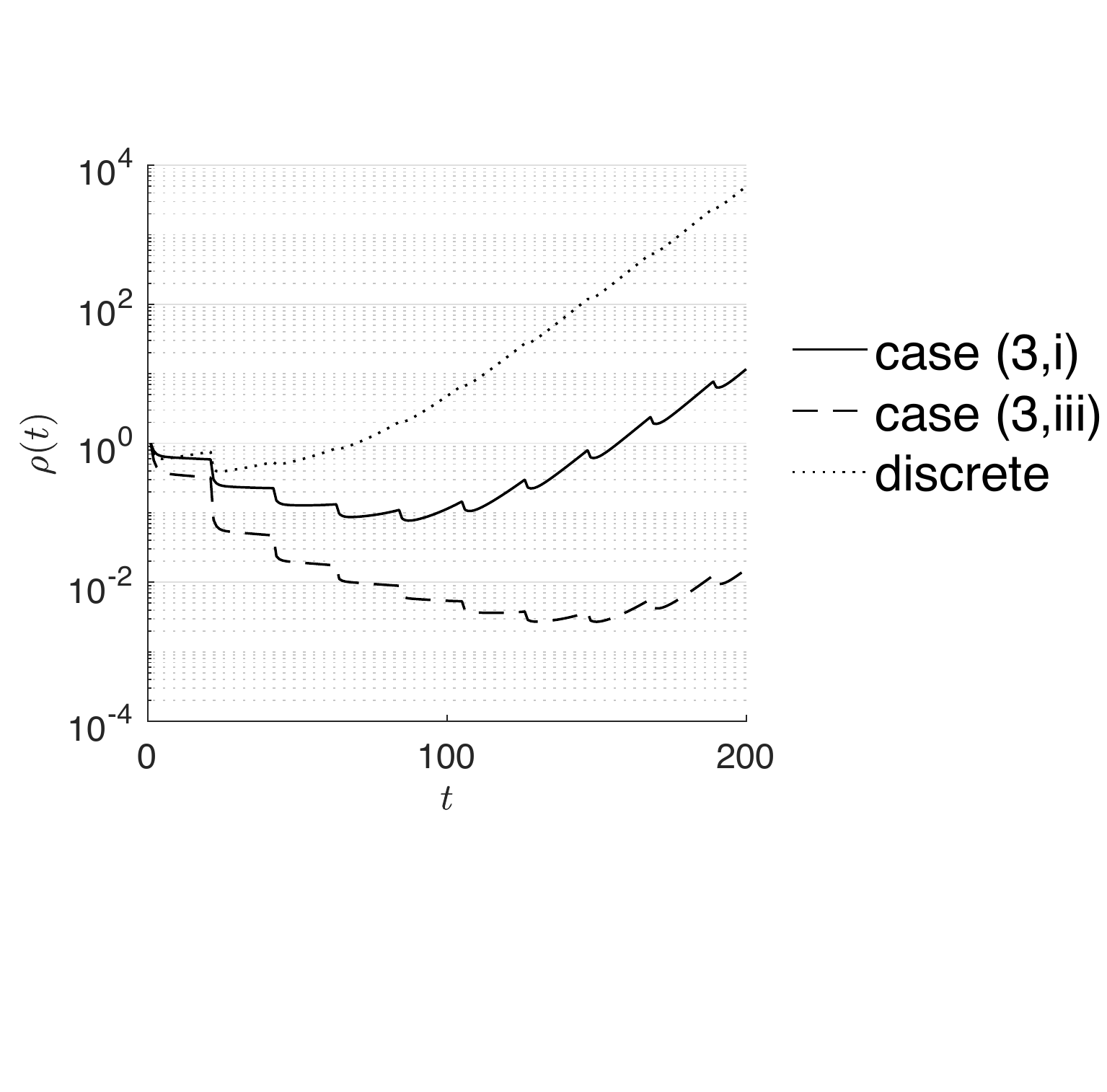}  
    }  
   \caption{Comparison of normalized total 
   number of cells $\log(\rho(t))$ at $t=200$ 
   when two drugs, either $[C_1,\, C_3]$ or $[C_3,\, C_6]$, 
   are applied in difference dosages $\omega_i c_i$.
   Top: three discrete levels of resistance \eqref{eq:GvrnGardner}.
   Bottom: the continuum model \eqref{eq:Gvrn0}--\eqref{eq:Gvrn1}.
   In the discrete model, the effects of drugs $C_1$, 
   $C_3$, and $C_6$ are binary depending on whether the drug is applied or not, 
   while the continuum models show gradual changes. 
   The figure on the right shows the results of using all six drugs, where 
   the tumor size significantly depends 
   on the choice of model (two orders of magnitude).
   } 
\label{fig:Grdnr_nTotClog}
\end{figure}

Figure \ref{fig:Grdnr_Pntp} compares the result of 
the discrete model \eqref{eq:GvrnGardner} and the continuum model
\eqref{eq:Gvrn0}--\eqref{eq:Gvrn1},
in particular with regards to the drug $C_2$. 
Shown is the cell distribution  
on the resistance trait space of drug $C_2$ 
in log scale\footnote{ 
$ n(t,\theta_2)  = \int_{\Gamma_i^c} n_P(t,\theta) + n_Q(t,\theta) d \theta_i^c$, where $\theta_i^c$ is the vector of $\theta$ except the $i$-th index $\theta_i$ and $\Gamma_i^c$ is its domain.
}, 
when using no drug, a single drug $c_2$, 
and all 6 drugs. 
Here, the continuum model is taken as the exponential kill model 
that can be classified as cases (iii) and (ii).
As expected from the shape of the uptake function  
in Figure \ref{fig:Grdnr_C}, 
the distribution in the $\theta_2$ trait space is 
concentrated 
at the boundary traits, similarly to the discrete model. 
However, 
the continuum model predict 
emerging cells with intermediate levels of resistance, 
and the degree of heterogeneity in the resistance level 
can be quantitatively computed.

Figure \ref{fig:Grdnr_nTotClog} 
compares the sensitivity of the tumor size with respect 
to the drug dosage 
between the continuum model \eqref{eq:Gvrn0}--\eqref{eq:Gvrn1} 
and the discrete model \eqref{eq:GvrnGardner}. 
For comparison, 
we plot the normalized 
total number of cells in log scale at time $t=200$ 
that is normalized by the mean. 
Here, two drugs are applied, 
either $(C_1,\, C_3)$ or $(C_3,\, C_6)$, with different 
weighted dosages $\omega_i c_i$, where $\omega_i = 0,\, 0.2,...,\, 1$. 
The results show that the tumor size $\rho(t)$ 
in the continuum 
model is more sensitive to the drug dosage, with variation of
a larger order of magnitude compared with
the results of the discrete model.
In addition, 
the effects of drugs $C_1$, $C_3$, and $C_6$ 
in the discrete model 
are binary depending on whether 
the drug is applied ($\omega_i \geq 0.2$) 
or not ($\omega_i = 0$). 
In contrast, the continuum model shows a gradual decay 
when increasing the dosage. 
Figure \ref{fig:Grdnr_nTotClog} also shows the 
total number of cells 
when all six drugs are applied. 
We observe that 
$\rho(t)$ significantly depends on the choice of model, 
as the tumor size varies
by two orders of magnitudes around $t=200$.

\begin{figure}[!htb]
    \centerline{ \footnotesize (a) $i=2$ \hspace{1.8cm} (b)  $i=4$ \hspace{1.8cm} (c)  $i=6$ \hspace{.6cm}  }
    \centerline{ 
    \includegraphics[width=11cm]{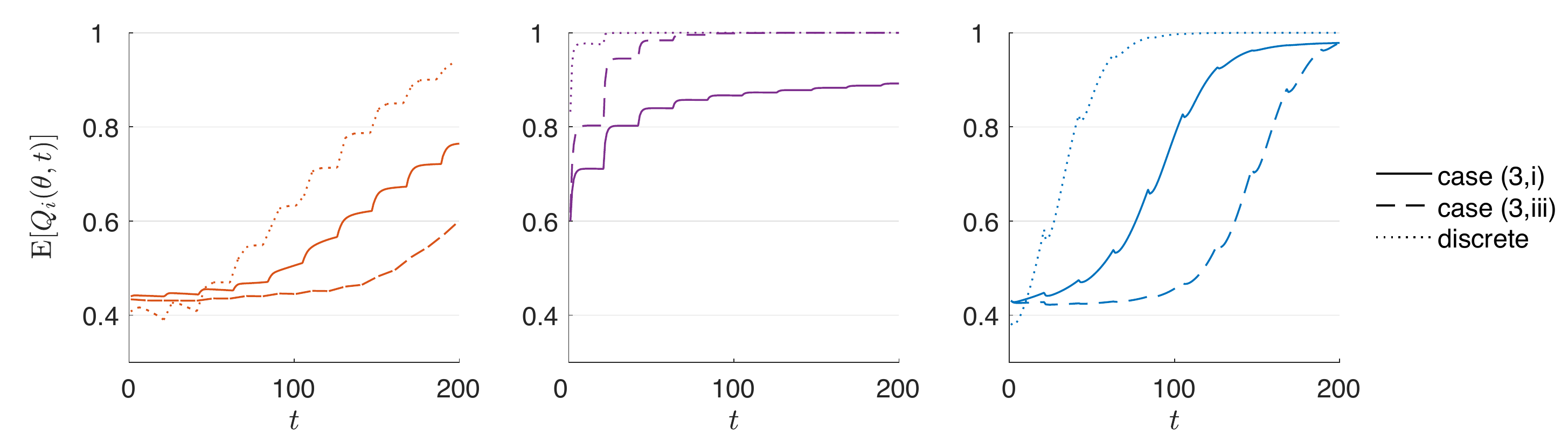} } 
   \caption{Comparison of the mean resistant trait E$[Q_i(t,\theta_i)]$ 
   to the $i$-th drug. 
   Each column corresponds to different types of drug:
   (a) CS cytotoxic, (b) nCS cytotoxic, and (c) cytostatic drug. 
   Using the discrete model \eqref{eq:GvrnGardner}, 
   the mean resistance level increases to $\theta_i = 1$ 
   in all drugs, i.e., the 
   cancer cell population is dominated by cells that are resistant to all six drugs. 
    In the continuum model \eqref{eq:Gvrn0}--\eqref{eq:Gvrn1}, resistance to CS cytotoxic drugs and
    cytostatic drug develops faster in case (i) compared with (iii), 
    while the resistance to nCS cytotoxic drug arises faster in case (iii). 
    } 
\label{fig:Grdnr_nTotC}
\end{figure}

Figure~\ref{fig:Grdnr_nTotC} compares 
the mean resistance level\footnote{
$E[Q_i(t,\theta_i)] \doteq \int \theta_i \, Q_i(t,\theta_i) d \theta_i $, where 
$Q_i(t,\theta) = \dfrac{ \int_{\Gamma_i^c} n_P(t,\theta) + n_Q(t,\theta) d \theta_i^c}{\rho(t)}$ and $\theta_i^c$ is the vector of $\theta$ except the $i$-th index $\theta_i$ and $\Gamma_i^c$ is its domain. 
} $E[Q_i(t,\theta_i)]$ 
up to $t=200$
when all 6 drugs are applied. 
While the mean resistance level in $\theta_i$ implies  
the dominating resistance 
to the $i$-th drug, 
we observe distinct results in different models. 
First, using the discrete model \eqref{eq:GvrnGardner}, 
the resistance level in each drug 
eventually converges to the most resistant cells 
$\theta_i = 1$.
This implies that 
the surviving cancer cells are 
only the ones that are fully resistant to all six drugs. 
However, the continuum model \eqref{eq:Gvrn0}--\eqref{eq:Gvrn1} 
shows a more gradual increase  
of resistance. 
Moreover, the resistance to nCS cytotoxic drugs 
develops more rapidly in case (iii) than in case (i). 
On the other hand, 
resistance to CS cytotoxic drugs and to cytostatic drugs 
is more sensitive to the drug application in case (i) that in case (iii). 
We finally comment that $E[Q_i(t,\theta_i)]$ 
shows similar dynamics when using drugs
with the same mechanism,
that is, the results with drugs
$C_1$, $C_3$, and $C_5$ are similar to 
$C_2$, $C_4$, and $C_6$, respectively. 

\begin{figure}[!htb]
    \centerline{ 
    \includegraphics[width=9cm]{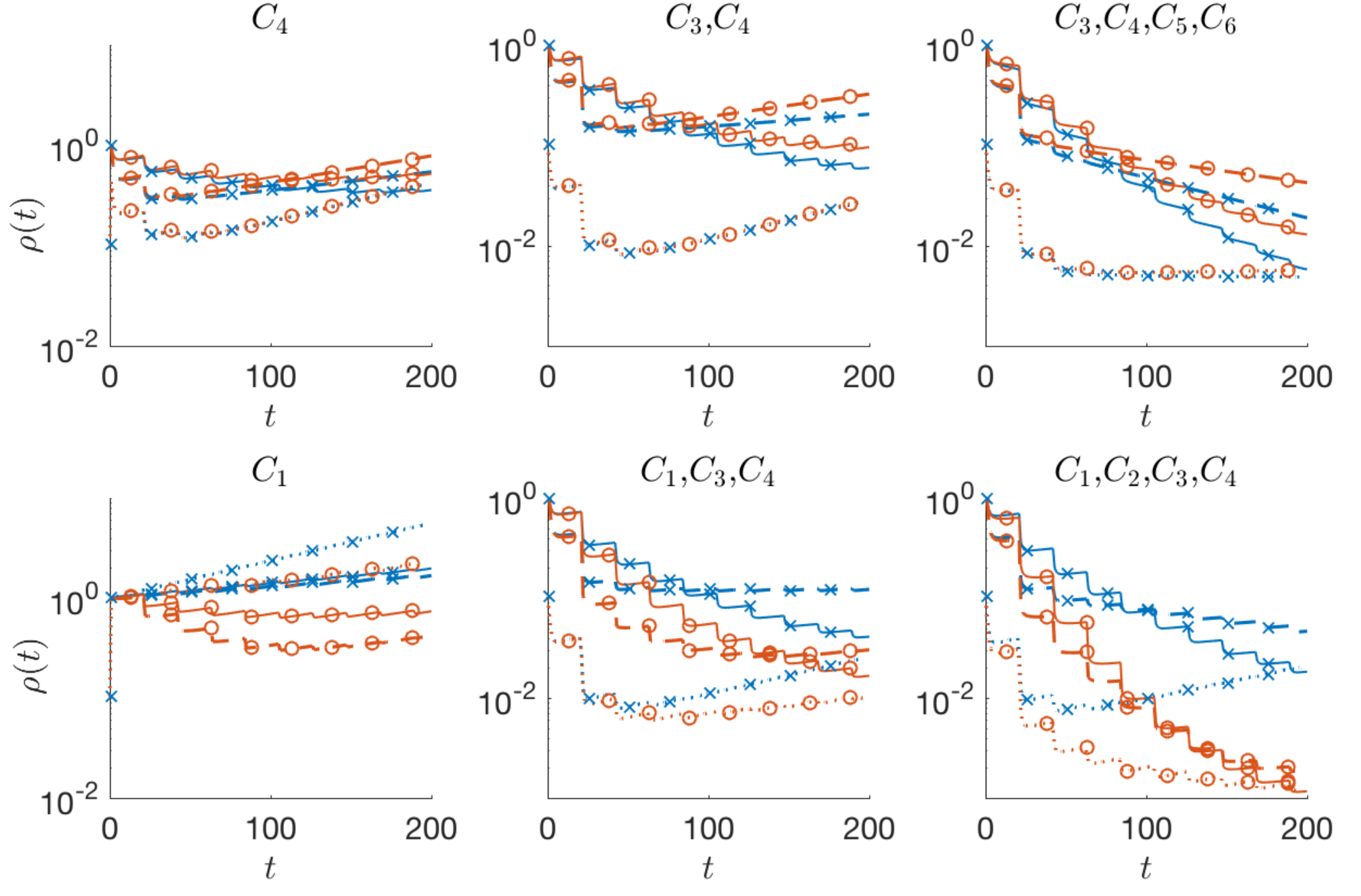} 
    \includegraphics[width=2.5cm]{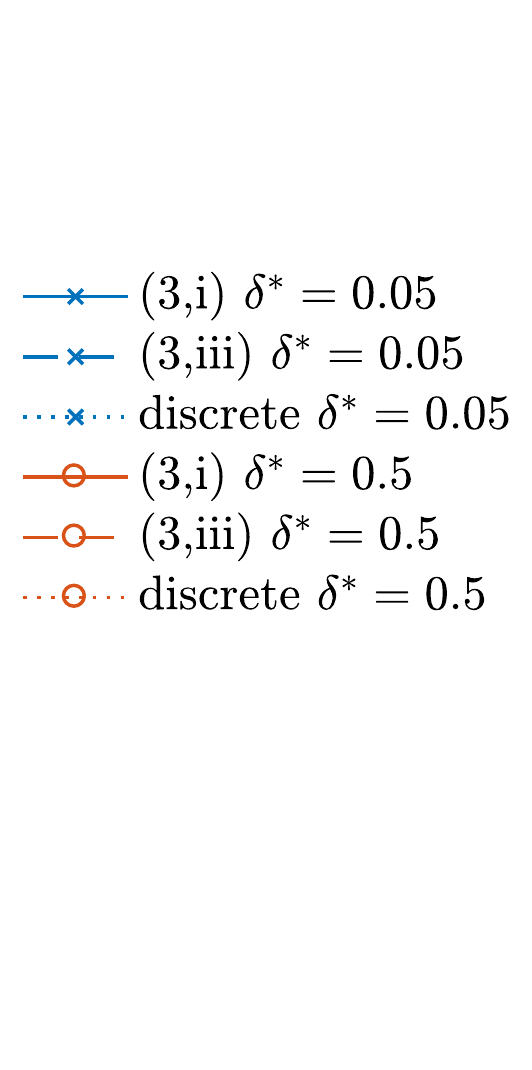} } 
   \caption{Total number of cells $\rho(t)$ using different drug combinations with either high or low proliferating index, that is, 
   $\delta^* = 0.5$ or $0.05$. 
   The drug combination that includes CS cytotoxic drugs ($C_1$ and $C_2$) are more effective in highly proliferating cells. 
   The drug effect of combinations without CS cytotoxic drugs 
   is independent of
   the proliferating index in the discrete model \eqref{eq:GvrnGardner}.
   In contrast, highly proliferating cancer cells show certain
   disadvantages in the continuum model \eqref{eq:Gvrn0}--\eqref{eq:Gvrn1}. } 
\label{fig:Grdnr2_nTotC}
\end{figure}

Gardner (2002) \cite{gardner2002} presents the effect of 
different drug combinations particularly 
to cancer cells with different proliferating 
proportions $\delta(t)$. 
Figure \ref{fig:Grdnr2_nTotC} shows 
simulations of 
the total number of tumor cells $\rho(t)$ 
with a highly proliferating index ($\delta^* =0.5$)
and a low proliferating index ($\delta^* = 0.05$). 
We demonstrate that 
the drug response function plays a key role in
determining the tumor growth dynamics 
using certain combination therapies 
that often involve 
the nCS cytotoxic drugs ($C_3$ and $C_4$). 
In general, 
the drug combinations that includes CS cytotoxic drugs 
($C_1$ and $C_2$) are more effective in 
highly proliferating tumors. 
In the discrete model \eqref{eq:GvrnGardner}, 
the drug combinations without the CS cytotoxic drugs show no difference.  
However, in the continuum model \eqref{eq:Gvrn0}--\eqref{eq:Gvrn1}, 
the highly proliferating cancer cells 
show disadvantage under drug combinations without 
CS cytotoxic drugs, 
which reveals a possible internal dependency between the drugs. 

We observe that 
the choice of continuum model is critical to 
the emerging drug response.
For an effective individually-tailored cancer modeling, 
these results stress the importance of identifying an appropriate model 
depending on the drug response of each individuals.

\section{Conclusion} \label{sec:Summary} 

In this paper we propose a mathematical model for multidrug resistance, 
assuming a continuous resistance phenotype space.
The multidrug resistance trait variable represents 
the level of resistance to various drugs including 
cell-cycle specific and nonspecific cytotoxic drugs,  
as well as cytostatic drugs.  
We classify the proliferation and drug uptake functions 
and identify 
the cases where the continuum model results in 
an intermediate maximal fitness resistance, 
i.e., the 
cases in which the continuum and discrete models are essentially 
different.
Thus, by observing the proliferation and drug effects, 
we can predict when 
the continuum models are different than the corresponding discrete
models.
We study the effect of epimutation 
on the cytotoxic and cytostatic resistance traits. 
In contrast to standard mutations that are associated with
an early relapse,
epimutations may either accelerate or delay the relapse time.
We demonstrate such effects on different 
continuum models, initial preexisting resistance ratios, 
and types of drugs. 

We use our approach to revising the works of Komarova and Wodarz
(2005) \cite{komarova2005} and the Gardner (2002) \cite{gardner2002}.
Following \cite{komarova2005}, 
we study the impact of the 
turnover rate on tumor growth and drug response.
We verify the effectiveness of 
a combination therapy with multiple cytotoxic drugs 
in low turnover tumors and also in high turnover tumors 
with a drug uptake function of case (i) under high drug dosages. 
Increasing the cytotoxic drug dosage 
delays the relapse 
in tumor that the drug uptake follows case (iii), 
but not in low turnover tumor with case (i),  
thus in particular in such cases, the dosage should be carefully chosen. 
Moreover, 
the choice of a drug uptake function is 
shown to have a higher impact
than the turnover rate 
under a logistic growth condition. 
These results provide new insights on the dynamics beyond what is
accessible by (and in certain cases even contradictory to) the
discrete-trait model of \cite{komarova2005}.

The second example we studied followed \cite{gardner2002} by
considering three different types of drugs:
cell cycle specific and nonspecific 
cytotoxic drugs, and cytostatic drugs.
We demonstrated that the size of the tumor 
is more sensitive to the drug dosage in 
the continuum models compared with the model of~\cite{gardner2002}.  
In addition, 
a drug combination without the 
cell cycle specific cytotoxic drug shows no disadvantage 
in highly proliferating tumors 
in the discrete model, 
which is not the case in the continuum models. 
We conclude that 
the dynamics of the cancer cell population 
including the time of relapse and the resistance profile
significantly depends on the choice of (continuum) models,
in addition to the turnover rate and the proliferation index.  
Thus, it is critical to select appropriate multidrug resistance models 
depending on the drug response of each individuals, 
to accomplish an effective individually-tailored cancer modeling framework 
and a corresponding optimal drug therapy.

Our future work includes deriving a continuum model 
from high-dimensional data that will be 
preprocessed with data analysis techniques. 
In addition, modeling the dependency structure 
of multiple drugs 
and investigating its effect on the resistance dynamics 
is another challenging topic. 
Finally, due to its dimensionality, 
simulation of multidrug resistance model 
requires developing an efficient numerical method 
that balances computational cost and 
accuracy.
This will be addressed with adaptive  
numerical methods that take advantage of 
the underlying low dimensional structure of the solution.

\section*{Acknowledgments}
The work of DL was supported in part by the National Science
Foundation under Grant Number DMS-1713109 and by
the Jayne Koskinas Ted Giovanis Foundation.

\appendix

\section{Parameters of simulation} \label{sec:appxA} 

The parameters for the simulation in section \ref{sec:Gardner} are taken from \cite{gardner2002} as following. 

\begin{itemize}

\item Maximum proliferation rate 
of highly proliferating cells is $\gamma = 1/30 $, 
and for less proliferating cells, 
it is $\gamma = 1/50$. 
In addition, 
reduced proliferation due to resistance is assumed that 
the cell cycle is delayed by approximately 20 hours 
\cite{STEEL1977,Baker1995,BEGG1997,PANETTA1997}.

\item Transfer rate from quiescent to proliferating cells: 
$p = 1/20 \dday^{-1}$ \cite{LORD2000, PAPAYANNOPOULOU2000, PANETTA1997}.  

\item Proliferating proportion: $0.05 \leq \delta^* \leq 0.5$ \cite{STEEL1977,Baker1995,BEGG1997,PANETTA1997}. 
We take $\delta^*=0.15$ unless otherwise stated. 

\item Necrosis rate of the quiescent cells:  
$D_Q = 1/100 \dday^{-1}$ \cite{LEITH1994}.  

\item $c_i(t) = \begin{cases} 
 \frac{\bc_i / d_i}{\lambda_i} \left( 1 - e^{-\lambda_i t} \right) + c_i^{prev}, \quad t \leq d_i \\ 
 \frac{\bc_i / d_i}{\lambda_i} e^{-\lambda_i t} \left( 1 - e^{-\lambda_i t} \right) 
 + c_i^{prev}, \quad t > d_i \\  
 \end{cases}, 
$
where $c_i^{prev}$ 
is the amount of drug built up from previous drug
applications and 
the parameters for drug administration are as follows \cite{CHABNER1993,GOLDENBERG1999,BCcancer}. 

\subitem -Periods of drug administration: 
$\lambda_i|_{i=1}^4 = 21$, 
$\lambda_5 = 1$, 
$\lambda_6 = 7$. 
\subitem -Duration of drug administration:   
$d_i|_{i=1}^4= 0.1$h, 
$d_5 = 2$h, 
$d_6 = 1/3$h. 
\subitem -Drug dosage scaled for $a_i = 2$ and $z_i = 1$:  
$\bc_1 = 5$, $\bc_2 = 0.005$, $\bc_3 = 0.0009$,  
$\bc_4 = 0.00012$, $\bc_5 = 0.01$, $\bc_6 = 0.01$.

\item Mutation rate: $w = 10^{-6}$ \cite{Coldman1998}.
\end{itemize}

\section*{References}

\bibliographystyle{elsarticle-num}
\bibliography{Tumor}

\end{document}